\documentstyle[eqsecnum,floats,aps,prb,epsfig,twocolumn]{revtex}

\begin{document}

\wideabs{

\title { 
 Hartree-Fock-LAPW method: using the full potential treatment for exchange } 
 
\author { 
A.V.~Nikolaev$^{a,b}$ and P.N.~Dyachkov$^c$ \\} 
\address{$^a$Department of Physics, University of Antwerp, UIA,
2610 Antwerpen, Belgium \\
$^b$Institute of Physical Chemistry of RAS, Moscow, Russia \\
$^c$Kurnakov Institute of General and Inorganic Chemistry of RAS, Moscow,
Russia
} 
 
\date{\today} 
 
\maketitle 
 
\begin{abstract}
We formulate a Hartree-Fock-LAPW method for electronic band 
structure calculations. The method is based on
the Hartree-Fock-Roothaan approach for  
solids with extended electron states and closed core shells where
the basis functions of itinerant electrons
are linear augmented plane waves.
All interactions within the restricted Hartree-Fock approach
are analyzed and in principle can be taken into
account. In particular, we have obtained the matrix elements
for the exchange interactions of extended states and the crystal
electric field effects.
In order to
calculate the matrix elements of exchange for extended states
we first introduce an auxiliary potential and then integrate it
with an effective charge density corresponding to
the electron exchange transition under consideration. The problem of
finding the auxiliary potential is solved by using the strategy
of the full potential LAPW approach, which is based on 
the general solution of periodic Poisson's equation.
Here we use an original technique for the general solution of 
periodic Poisson's equation and multipole expansions of electron densities.
We apply the technique to obtain 
periodic potentials of the face centered cubic lattice
and discuss its accuracy and convergence in comparison with
other methods.
\end{abstract} 
 
%

}

\narrowtext

\section {Introduction} 
\label{sec:int} 

In contrast to successful applications of
the Hartree-Fock (HF) method to molecules, which have become a routine
procedure, Hartree-Fock calculations of
electron band structure of solids are relatively rare.
The electronic band structure calculations mainly rely on
the density functional approach \cite{DFT,Arg} with
the local density approximation (LDA) or other non-local exchange 
approximations \cite{GGA}
which try to mimic the Fock exchange for conduction
electrons. 
As known, LD approximation is a product of the Hartree-Fock
theory of the free electron gas and as such it inherits its 
shortcomings. \cite{Nes}
The LD approximation being universal and simple from one side, is
structureless and leads to an unphysical self-interaction, \cite{SIC}
from the other.
In addition, LDA is an uncontrolled
approximation, and there is no straightforward path to further
improvements in its accuracy. \cite{Arg}
Therefore, it is highly desirable to combine the Hartree-Fock
approach with methods based on plane wave basis sets 
which traditionally play an important role in the field of
electron band structure calculations.
Although a variety of computational procedures on periodic HF
method has been reported in the literature, 
\cite{HFR,Pis,CRY,Sau,Dov,Kud,Andre,Vis}
all of them use Gaussian-type orbitals (GTO). 
In this paper we present a merger of the Hartree-Fock approach with
the linear augmented plane wave \cite{And,Koe,Sin} (LAPW) method, 
which constitutes a new Hartree-Fock-LAPW procedure for electron 
band structure calculations.
The consideration of the LAPW basis is important since the method
has been extensively tested and it is traditionally associated
with electronic band structure calculations.

At the center of the Hartree-Fock-Roothaan approach  there are
calculations of the direct Coulomb and the exchange matrix elements \cite{Roo}
and in the present work we consider thoroughly
all cases, including the delocalized and
localized states. The advantage is that all expressions for the
matrix elements can be found in one source ({\it i.e.} this paper),
but on the other hand, this has led to numerous formulas and 
their extensive derivations.
While GTOs are popular in calculations
of molecular systems and have the advantage of giving easily
integrable polycentric functions, we lose this benefit working
with the LAPW basis set. There,
we have to rely on a different strategy of calculations.
Our choice here is the full potential LAPW (FP-LAPW)
treatment, \cite{Wei,Rud} which is based on the procedure of 
finding the general solution for a periodic Poisson's equation.
For Poisson's equation  we use a new original technique (Sec.\ II).
It employs the Ewald method \cite{Ewa,Ewa1} 
for the potential of the monopole terms and an exact expansion
 in Fourier series in the interstitial region,
with a multipole series expansion of the potential inside the spheres. 

A principal new element of our work is to use the strategy of
the FP-LAPW method for
the calculation of exchange matrix elements between
itinerant electrons.
Indeed, an exchange matrix element 
$\langle a b| V^{Coul} |b a \rangle$
($a$ and $b$ here represent electronic states with the same spin
component) can be estimated as a Coulomb self-interaction energy
of the charge density distribution $\psi_b^*(\vec{R}) \psi_a(\vec{R})$, 
which appears as a result of the electron
transition $a \rightarrow b$.
The procedure implies (i) an introduction of an auxiliary potential
of the charge density $\psi_b^*(\vec{R}) \psi_a(\vec{R})$ and
(ii) the integration of the potential with the charge density.
Such treatment is rigorous and
ensures that {\it all contributions including
the long range ones} (like the exchange Ewald sums) are carefully considered.
It is worth to notice that a simple truncation
of the exchange series of other methods
excludes such long range effects. 
 
Here
we also consider crystal electric field (CEF) effects for the closed 
shell core electrons.
Usually in the full potential LAPW treatment the 
core electrons are taken in the spherical
approximation and the CEF effects are ignored. \cite{Sin} 
While the crystal field splittings are small
it may be still necessary to include these effects 
for precise calculations of solids where
one compares the stability of different crystalline phases where
the energy difference is of the order of 100~K. 
The present consideration is a generalization of 
previous works on CEF effects. \cite{Nik1,Nik2,Nik3,Hut}

For clearness and simplicity in the following we consider 
only one atom per unit cell. However, the method can
be easily generalized for the case of a few atoms. 
In this paper we consider the core states being 
completely confined inside the spheres.
Various improvements of the method in this respect are also
possible, \cite{Sin} but they are not of fundamental
importance and will not be concerned here.

The paper is organized as follows. We start with the
general solution of periodic Poisson's equation, Sec.\ II.
In Sec.\ III we illustrate how this technique works for the
case of a face centered cubic (fcc) crystal. We
compare the convergence and accuracy of
different approaches.
Then we discuss the Hartree-Fock-Roothaan method
for solids with extended states, Sec.\ IV, and
 calculate the direct Coulomb matrix elements
for LAPW basis functions. 
In Sec.\ V we present calculations
of the matrix elements of exchange,
while Sec.\ VI is reserved for our conclusions.

\section {General solution of periodic Poisson's equation} 
\label{sec:i} 

The general solution of Poisson's equation $V^{Coul}$ for 
a periodic charge density
was demanded by necessity to improve the ``muffin-tin" (MT)
approximation of LAPW method. \cite{And,Koe,Sin}
LAPW electron band structure calculation procedure with the 
general potential
$V^{tot}(\vec{R})=V^{Coul}(\vec{R})+V^{exc}(\vec{R})$ 
is referred to as the full potential LAPW (FP-LAPW) method \cite{Wim} since in
principle no shape approximation (except LDA) for the potential is assumed. 
First full potential calculations have been reported in 
Refs.\ \onlinecite{Ham,Wim}, and since then they are often used
to investigate the electron band structure of solids. 
These FPLAPW calculations are based on the general solution of 
Poisson's equation given by Weinert in Ref.\ \onlinecite{Wei}. 
The basic idea which has been proposed already in Ref.~\onlinecite{Rud}
includes two steps:
1) to obtain the potential in the interstices and then
2) to solve the boundary value problem inside a sphere.
The point 1) in Ref.~\onlinecite{Wei} 
represents an alternative to the Ewald method \cite{Ewa,Ewa1} 
where the true charge inside the spheres is replaced with
a pseudo-charge-density of the same multipole moments. 
In the present paper we propose a new technique which uses the Ewald
expansion for the monopoles ($l=0$) and the exact Fourier expansions
for the higher multipoles ($l>0$) in the interstitial region. 
The procedure is
straightforward and besides a truncation of
the multipole series, it
contains only one cut-off parameter
for the Ewald expansion. The strategy consists of the same two
steps as outlined above. \cite{Rud,Wei}

\subsection {Dual and equivalent representations of charge density}

We start with the description of charge density
in a crystal.
For a periodic crystalline structure we introduce $\vec{n}$
to label the sites and the direct lattice vector 
$\vec{X}(\vec{n})$ which specifies the centers of atoms (spheres).
The position vector $\vec{R}$ inside a sphere $S(\vec{n})$
is given by
\begin{equation}
 \vec{R}(\vec{n})=\vec{X}(\vec{n})+\vec{r}(\vec{n}) .
\label{2.1}  
\end{equation}
In spherical coordinates $\vec{r}(\vec{n})=(r(\vec{n}),\hat{r}(\vec{n}))$,
where $\hat{r}\equiv \Omega=(\theta,\phi)$.
For a Bravais lattice of $\vec{X}(\vec{n})$ we construct the
reciprocal lattice of vectors $\vec{K}$. 
It is convenient to
partition the space into
the region $S$ inside the non-overlapping spheres of radius $R$
and the interstitial
region $I$, Fig.~1.
%
\begin{figure} 
\vspace{-0.6cm}
\centerline{ 
\epsfig{file=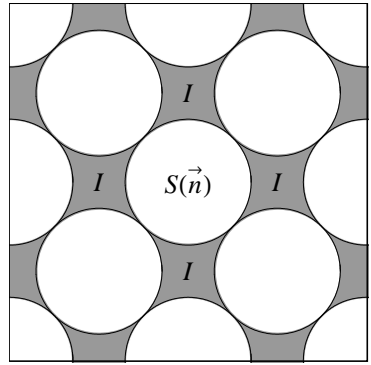,width=0.47\textwidth} 
} 
\vspace{-5.3cm}
\caption{
The region inside the spheres $S$ (white) and the interstitial
region $I$ (gray).
} 
\label{fig1} 
\end{figure} 
We expand the charge density in Fourier series in region $I$
and in multipolar terms in region $S$ (dual representation \cite{Ely,Wim}):
\begin{mathletters}
\begin{eqnarray}
& & \left. e\rho(\vec{R}) \right|_{\vec{R} \in I} =  
  e \sum_{\vec{K}} \rho_I(\vec{K})\, e^{i \vec{K} \vec{R}} , 
  \label{2.2a}     \\
& & \left. e\rho(\vec{R}) \right|_{\vec{R} \in S} =  
  e \left( \sum_{\Lambda} \rho_{\Lambda}(r)\, S_{\Lambda}(\hat{r})-Z\, 
  \frac{\delta(r)}{\sqrt{4\pi} r^2} S_{0} \right) .
 \label{2.2b}  
\end{eqnarray}
\end{mathletters}
Angular functions $S_{\Lambda}$ are symmetry adapted functions \cite{Bra} 
(SAFs) which are linear combinations of spherical harmonics
(see Appendix A for details and definitions).
The total density is an invariant of the space group under consideration.
Therefore, in the expansion (\ref{2.2b}) we use only SAFs $S_{\Lambda}$
of the full (or unit) $A_{1g}$ symmetry. 
Finally, $\delta(r)$ in (\ref{2.2b}) is the one dimensional 
(radial) delta-function,
$Z$ is the atomic number and $e<0$ is the charge
of electron. In the following we will use atomic units and the
charge density is understood in units $e=-1$.
The density expansion (\ref{2.2a},b) called also 
the dual representation \cite{Ely,Wim}
is characteristic of the LAPW method. \cite{And,Koe,Sin} 
In fact, the dual representation 
is natural because in the interstitial region
electron density is smooth whereas large charge oscillations near 
nuclei require a multipole analysis.

Using the linearity of Poisson's equation we can extend smoothly the
interstitial Fourier charge representation inside the spheres
and subtract it out. We obtain
\begin{mathletters}
\begin{eqnarray}
 & & \rho(\vec{R})=\rho_I(\vec{R})+\sum_{S(\vec{n})} 
 \rho'_{S(\vec{n})}(\vec{R}) ,
  \label{2.1a} \\  
 & & \rho'_{S(\vec{n})}(\vec{R})= 
 (\rho_{S(\vec{n})}(\vec{R})-\rho_I(\vec{R}))\, 
 \Theta(\vec{R} \in S(\vec{n})),
  \label{2.1b} 
\end{eqnarray}
\end{mathletters}
where $\rho_I(\vec{R})$ is defined by (\ref{2.2a})
 over the whole space, 
$\rho'_{S(\vec{n})}$ is the renormalized charge density inside 
a sphere $S(\vec{n})$ and $\Theta$ is the unit step function.
Using Eq.\ (\ref{2.1}) and the plane wave expansion in SAFs
centered at the site $\vec{n}$,
\begin{eqnarray}
 e^{i \vec{K} \vec{R}} =  e^{i \vec{K} \vec{X}(\vec{n})}
 4\pi \sum_{\Lambda} i^l j_l(Kr(\vec{n})) 
 S_{\Lambda}(\hat{K}) S_{\Lambda}(\hat{r}(\vec{n})),
  \label{2.2}  
\end{eqnarray}
we rewrite the multipolar density inside the sphere as
\begin{mathletters}
\begin{eqnarray}
 & & \rho'_0(r)=\rho_0(r)-\sqrt{4 \pi} \rho_I(\vec{K}=0) 
 -Z\, \frac{\delta(r)}{\sqrt{4\pi} r^2} , \label{2.3a} \\
 & & \rho'_{\Lambda}(r)=\rho_{\Lambda}(r) -4\pi i^l
 {\sum_{\vec{K} \neq 0}}' j_l(Kr) S_{\Lambda}(\hat{K}) \rho_I(\vec{K}) .  
 \label{2.3b}
\end{eqnarray}
\end{mathletters}
We shall refer to (\ref{2.1a},b) with the multipole densities 
(\ref{2.3a},b) as to the equivalent representation of the charge density.
We will use it every time when we need to obtain the potential 
in the interstitial region
because the Fourier series (\ref{2.2a}) is now extended
to the whole crystal, that has some important advantages (see next subsection).

In principle, in Eq.\ (\ref{2.2a})
there is a term with $\vec{K}=0$ that corresponds to the homogeneous
electron density distribution. However, it will not contribute
to the resulting potential because it cancels with the positive
homogeneous density distribution arising from nuclei (see also next subsection).

\subsection {The solution in the interstitial region}

The potential in the interstices has three components.
The first contribution arises from the density $\rho_I(\vec{R})$ given by
the Fourier series (\ref{2.2a}) throughout the whole crystal. The second and
third terms are due to the monopole ($l=0$) and the high multipole
($l \ge 1$) moments of electron density inside the spheres.
(We recall that here we are dealing with the equivalent charge density
given by Eqs.\ (\ref{2.3a},b).)

From the Fourier series of $\rho_I(\vec{R})$ we find
\begin{eqnarray}
 V_I(\vec{R})={\sum_{\vec{K} \neq 0}}' 
 V^I(\vec{K})\, e^{i \vec{K} \vec{R}} ,
  \label{2.5}  
\end{eqnarray}
where
\begin{eqnarray}
 V^I(\vec{K})=\frac{4\pi}{K^2} \rho_I(\vec{K}) .
  \label{2.6}  
\end{eqnarray}
The easiness of the last expression owes to the fact that we
extend the Fourier expansion of $\rho_I(\vec{R})$ to the whole
crystal. A complication is that the density inside a sphere
is renormalized according to Eqs.\ (\ref{2.3a},b). 
Next we consider a sphere $S(\vec{n})$ and for a moment
we will not include to the potential the contribution 
due to the charges outside the sphere. 
Thus we introduce a single sphere potential
${\cal U}_S(\vec{r})={\cal U}_{S(\vec{n})}(\vec{r}(\vec{n}))$.
Using the one-site expansion \cite{Jac} in SAFs
\begin{eqnarray}
 \frac{1}{|\vec{r}(\vec{n})-\vec{r}'(\vec{n})|}= \sum_{\Lambda}
 v_{\Lambda \Lambda}(\vec{n}=\vec{n}',r,r')
 S_{\Lambda}(\hat{r}') S_{\Lambda}(\hat{r}),
  \label{2.7}  
\end{eqnarray}
where 
\begin{eqnarray}
 v_{\Lambda \Lambda'}(\vec{n}=\vec{n}',r,r')
 & = & \int d\Omega \int d \Omega'
 \frac{S_{\Lambda}(\Omega) S_{\Lambda'}(\Omega')}
 {|\vec{r}(\vec{n})-\vec{r}'(\vec{n})|} \nonumber \\
 & = & \delta_{\Lambda \Lambda'} \frac{4\pi}{2l+1} \frac{r_<^l}{r_>^{l+1}} ,
 \label{3.16d}
\end{eqnarray}
and $r_<$ ($r_>$) is the smaller (larger) of $r$ and $r'$,
we obtain the solution inside
the sphere ($0<r \leq R$),
\begin{eqnarray}
 {\cal U}_{S}(\vec{r}) &=& \frac{Q_0(r)}{r}+Q'_0(r) \nonumber \\
 &+& {\sum_{\Lambda \neq 0}}' \frac{4\pi}{2l+1}
 \left( \frac{Q_{\Lambda}(r)}{r^{l+1}}+r^l 
 Q'_{\Lambda}(r) \right) S_{\Lambda}(\hat{r}) .
  \label{2.8}  
\end{eqnarray}
Here we have introduced the functions
\begin{mathletters}
\begin{eqnarray}
 & & Q_0(r)= 
 q_0(r)-\frac{4\pi r^3}{3} \rho_I(\vec{K}=0) ,
  \label{2.9a} \\
 & & Q'_0(r)=q'_0(r)
 -2 \pi (R^2-r^2)\, \rho_I(\vec{K}=0) ,
 \label{2.9b} 
\end{eqnarray}
\end{mathletters}
with
\begin{mathletters}
\begin{eqnarray}
 & & q_0(r)= 
 \sqrt{4 \pi} \int_0^r \rho_0(r')\, {r'}^2 dr'-Z ,
  \label{2.10a} \\
 & & q'_0(r)=\sqrt{4 \pi} \int_r^R \rho_0(r')\, r' dr' , 
 \label{2.10b} 
\end{eqnarray}
\end{mathletters}
for the spherically symmetric term $\Lambda=l=0$, and
\begin{mathletters}
\begin{eqnarray}
 & & Q_{\Lambda}(r)= 
  \int_0^r \rho'_{\Lambda}(r')\, {r'}^{\, l+2} dr' ,
  \label{2.11a} \\
 & & Q'_{\Lambda}(r)= \int_r^R \rho'_{\Lambda}(r')\, {r'}^{\,1-l} dr' ,
 \label{2.11b} 
\end{eqnarray}
\end{mathletters}
for the other multipoles, $\Lambda \neq 0$.
Taking into account the explicit expression for $\rho'(r)$, Eq.\ (\ref{2.3b}),
we rewrite (\ref{2.11a},b) as
\begin{mathletters}
\begin{eqnarray}
   Q_{\Lambda}(r) &=& 
   q_{\Lambda}(r) 
   - 4\pi i^l r^{l+2} {\sum_{\vec{K} \neq 0}}' S_{\Lambda}(\hat{K}) 
  \rho_I(\vec{K}) \frac{j_{l+1}(Kr)}{K} ,  \nonumber \\
  & & \label{2.12a} \\
  Q'_{\Lambda}(r) &=&  q'_{\Lambda}(r)
   + 4\pi i^l {\sum_{\vec{K} \neq 0}}' S_{\Lambda}(\hat{K}) \rho_I(\vec{K}) 
   \nonumber \\
   & \times & 
   \frac{1}{K}
  \left( \frac{j_{l-1}(KR)}{R^{l-1}}-
  \frac{j_{l-1}(Kr)}{r^{l-1}} \right) ,
 \label{2.12b} 
\end{eqnarray}
where
\begin{eqnarray}
   q_{\Lambda}(r) &=& 
  \int_0^r \rho_{\Lambda}(r')\, {r'}^{\, l+2} dr' ,
  \label{2.12c} \\
  q'_{\Lambda}(r) &=& \int_r^R \rho_{\Lambda}(r')\, {r'}^{\,1-l} dr' . 
  \label{2.12d} 
\end{eqnarray}
\end{mathletters}
In order to obtain Eq.\ (\ref{2.12a},b) we have performed integration 
of spherical Bessel functions using
the properties 10.1.23, 10.1.24 of
Ref.\ \onlinecite{Abr}. 

For the potential ${\cal U}_S(\vec{r})$ outside 
the sphere $S(\vec{n})$ ($r \ge R$) we have
\begin{eqnarray}
 {\cal U}_S(\vec{r}) = \frac{Q_0(R)}{r} 
 +{\sum_{\Lambda \neq 0}}' \frac{4\pi}{2l+1}
  \frac{Q_{\Lambda}(R)}{r^{l+1}}  S_{\Lambda}(\hat{r}) . 
  \label{2.12'}  
\end{eqnarray}
Now we are ready to compute the potential of the periodic 
arrangements of the spheres,  
\begin{eqnarray}
  V_S(\vec{R}) = \sum_{\vec{n}} {\cal U}_S(\vec{R}-\vec{X}(\vec{n})) .
  \label{2.4} 
\end{eqnarray}
Since we seek for
 the solution in the interstitial region
it is desirable to expand it in a Fourier series,
in the spirit of the dual representation,
\begin{mathletters}
\begin{eqnarray}
 V_S(\vec{R})={\sum_{\vec{K} \neq 0}}' 
 V^S(\vec{K})\, e^{i \vec{K} \vec{R}} .
  \label{2.14a}   
\end{eqnarray}
We distinguish two contributions, $V^S_0$ and $V^S_M$, from the
spherically symmetric charge distribution and from the other
higher charge multipoles, respectively:
\begin{eqnarray}
 V^S(\vec{K})=V^S_0(\vec{K})+V^S_M(\vec{K}) . \label{2.14b} 
\end{eqnarray}
\end{mathletters}
The monopole term ($l=0$) $V^S_0$ can not be computed directly
and requires a special treatment. Here we employ 
the Ewald method \cite{Ewa,Ewa1}
and obtain for the Fourier coefficients
\begin{eqnarray}
 V^S_0(\vec{K}) =\frac{\pi}{v} \frac{4}{K^2}\,e^{-K^2/4G^2} Q_0(R) ,
  \label{2.15}  
\end{eqnarray}
where $G$ is the Ewald cut-off parameter 
and $v$ is the unit cell volume, $v=V/N$. The improved Ewald
technique requires an additional summation in real space, 
see Ref.~\onlinecite{Ewa1}. Since we are concerned with the
Fourier expansion in the interstitial region we can always
take $G$ large enough and neglect the summation in real
space. (At large $G$ the dimensionless 
parameter $c=GR \gg 1$ and the complementary
error function decreases exponentially.)
Then the Ewald expansion with the Fourier coefficients (\ref{2.15})
converges absolutely in the interstitial
region. 

In the Ewald expansion, Eq.\ (\ref{2.15}), and in the Fourier decomposition 
(\ref{2.5}) we have omitted the terms
with $\vec{K}=0$ which correspond to a homogeneous charge distribution
of the electrons and the nuclei.
Both these terms, considered separately, diverge. Their sum, however,
as a consequence of electroneutrality gives the zero value of
the charge density and therefore can be discarded.

The high multipole terms ($l \geq 1$) converge fast. 
Their Fourier coefficients are given by
\begin{eqnarray}
  V^S_M(\vec{K}) =\frac{1}{v} \int_{cell}
 e^{-i \vec{K} \vec{R}}\, V_{S_M}(\vec{R})\, d^3R .
  \label{2.15a}
\end{eqnarray}
Here the integration is over any primitive unit cell of the crystal,
and $V_{S_M}(\vec{R})$ is the potential (\ref{2.4}) where only
the high multipoles ($l \ge 1$) are retained in ${\cal U}_S$.
Since the result of integration (\ref{2.15a}) does not depend on
the choice of unit cell, it follows  \cite{Ewa1} 
that the coefficients $V^S_M(\vec{K})$ can be rewritten as
\begin{eqnarray}
 V^S_M(\vec{K}) =\frac{1}{v} \int_{crystal} e^{-i \vec{K} \vec{R}}\,
 {\cal U}_{S_M}(\vec{R}) \, d^3R .
  \label{2.15b}  
\end{eqnarray}
Here the integration is taken over the whole crystal and
${\cal U}_{S_M}(\vec{R})$ is given by the 
last (high multipole) term on the right hand
sides of (\ref{2.8}) and (\ref{2.12'}) for $0<|\vec{R}| \leq R$ and
$|\vec{R}| > R$, respectively.
Using Eq.\ 10.1.24 of Ref.~\onlinecite{Abr} we find
\begin{mathletters}
\begin{eqnarray}
  V^S_M(\vec{K}) &=& \frac{(4\pi)^2}{v}  {\sum_{\Lambda \neq 0} }'
 \frac{ (-i)^l}{2l+1} S_{\Lambda}(\hat{K}) \nonumber \\
 & & \times
 \left(  A_{\Lambda}(K) + Q_{\Lambda}(R) \frac{j_{l-1}(KR)}{K R^{l-1}} \right) ,
 \label{2.16a}  
\end{eqnarray}
where
\begin{eqnarray}
 A_{\Lambda}(K)= \int_0^R \left( \frac{Q_{\Lambda}(r)}{r^{l+1}} + 
 r^l Q'_{\Lambda}(r) \right) j_l(Kr) r^2 dr .
 \label{2.16b}  
\end{eqnarray}
\end{mathletters}
(Note that at $r \rightarrow 0$, $Q_{\Lambda} \rightarrow r^{l+2}$,
$Q'_{\Lambda} \rightarrow r^{1-l}$,
$j_l(Kr) \rightarrow r^l$ and the integral (\ref{2.16b})
 does not diverge.)  
 
Collecting all contributions together, Eqs. (\ref{2.6}), (\ref{2.15}), 
(\ref{2.16a}),
we obtain for the Coulomb potential in the interstitial region,
\begin{mathletters}
\begin{eqnarray}
 V(\vec{R} \in I)={\sum_{\vec{K} \neq 0}}' 
 V(\vec{K})\, e^{i \vec{K} \vec{R}} ,
 \label{2.17a}  
\end{eqnarray}
where
\begin{eqnarray}
 V(\vec{K})=V^I(\vec{K})+V^S_0(\vec{K})+V^S_M(\vec{K}) .
 \label{2.17b}  
\end{eqnarray}
\end{mathletters}

\subsection {The solution inside the spheres}

The potential in the form (\ref{2.17a},b) can not be extended directly
to the region $S$ inside the spheres. The main reason is that it requires 
an infinite number of
plane waves to describe the Coulomb nucleus singularity. 
In order to obtain a well-behaved solution
we expand the total potential $V(\vec{R})$ 
inside a sphere $S(\vec{n})$ ($r(\vec{n}) \le R$) 
in a multipole series \cite{Rud,Wei}
\begin{eqnarray}
     V(\vec{r}(\vec{n}))=V_0(r)+{\sum_{\Lambda \neq 0}}'
     V_{\Lambda}(r)\, S_{\Lambda}(\hat{r}) ,
 \label{2.18} 
\end{eqnarray}
The potential $V(\vec{r}(\vec{n}))$ is defined uniquely by
boundary conditions. \cite{Jac}
These are of the Dirichlet kind here, since by using
the expansion (\ref{2.17a}) for $V(\vec{R} \in I)$ on the sphere surface, 
we arrive at
\begin{mathletters}
\begin{eqnarray}
  & & V_0(R)=  {\sum_{\vec{K} \neq 0}}'
 j_0(KR) V(\vec{K}) ,
 \label{2.19a}      \\
  & & V_{\Lambda}(R)=4\pi i^l {\sum_{\vec{K} \neq 0}}'
 j_l(KR) S_{\Lambda}(\hat{K}) V(\vec{K}) .
 \label{2.19b}
\end{eqnarray}
\end{mathletters}
Now inside the sphere $S(\vec{n})$ we have a Dirichlet problem. \cite{Wei,Jac}
We solve it by considering the spherical Green's function expansion, \cite{Jac}
and rewrite the solution (Eqs.\ (3.126) and (3.114) of 
Ref.~\onlinecite{Jac}) in the following form:
\begin{eqnarray}
  V(\vec{r}(\vec{n}))=U_S(\vec{r}(\vec{n}))
  +V^{out}_S(\vec{r}(\vec{n})),
  \label{2.20}
\end{eqnarray}
The potential $U_S(\vec{r}(\vec{n}))$ is induced by
the charge distribution {\it inside the sphere} $S(\vec{n})$. Clearly, it is 
independent of site $\vec{n}$ and we find 
\begin{eqnarray}
 U_S(\vec{r}) &=& \frac{q_0(r)}{r}+q'_0(r) \nonumber \\
 &+& {\sum_{\Lambda \neq 0}}' \frac{4\pi}{2l+1}
 \left( \frac{q_{\Lambda}(r)}{r^{l+1}}+r^l 
 q'_{\Lambda}(r) \right) S_{\Lambda}(\hat{r}) .
  \label{2.21}  
\end{eqnarray}
Notice that the potential $U_S$ differs from 
${\cal U}_S$, Eq.\ (\ref{2.12'}). In (\ref{2.21}) we use
$q_{\Lambda}$, Eq.\ (\ref{2.12c}), and $q'_{\Lambda}$, Eq.\ (\ref{2.12d})
(not $Q_{\Lambda}$ and $Q'_{\Lambda}$!).
$q_{\Lambda}$ and $q'_{\Lambda}$ correspond to
the initial charge density inside the sphere, Eq.\ (\ref{2.2b}), rather
than the equivalent one, Eqs.\ (\ref{2.3a},b).
$V^{out}_S(\vec{r})$ is the potential due to the charge
situated {\it outside the sphere} $S(\vec{n})$,
\begin{eqnarray}
  V^{out}_S(\vec{r})=V^{out}_{0}(R) + 
  {\sum_{\Lambda \neq 0}}' V^{out}_{\Lambda}(R) \left( \frac{r}{R} \right)^l
  S_{\Lambda}(\hat{r}) .
  \label{2.22}
\end{eqnarray}
Here the constants $V_0^{out} \equiv V_0^{out}(R)$ and 
$V_{\Lambda}^{out} \equiv V_{\Lambda}^{out}(R)$
are given by
\begin{mathletters}
\begin{eqnarray}
 V^{out}_{0} &=&   V_0-\frac{q_0(R)}{R} ,
   \label{2.23a} \\
 V^{out}_{\Lambda} &=& V_{\Lambda}-\frac{4\pi}{2l+1}
 \frac{q_{\Lambda}(R)}{R^{l+1}} .
  \label{2.23b}
\end{eqnarray}
\end{mathletters}
In the following we drop the argument $R$ in the
boundary values $V_0 \equiv V_0(R)$ and $V_{\Lambda} \equiv V_{\Lambda}(R)$.
Notice that at $\vec{r}=0$ the potential 
due to the charges outside the sphere $S(\vec{n})$ is given only by 
$V^{out}_0$ since the other contributions ($l \neq 0$) reduce to zero. 
  
Recalling that the Fourier components of the
interstitial potential (\ref{2.17a},b) consists of three parts, 
we distinguish the same three contributions
on the sphere surface for $V_0$ and $V_{\Lambda}$ in Eqs.\ (\ref{2.19a},b):
\begin{eqnarray}
  V_{\Lambda}=V_{\Lambda}^I+V_{\Lambda}^{S_0} 
  +V_{\Lambda}^{S_M} . 
  \label{2.27}
\end{eqnarray}
Here $V_{\Lambda}^I$, $V_{\Lambda}^{S_0}$ and $V_{\Lambda}^{S_M}$
are obtained by using Eqs.\ (\ref{2.19a},b) where instead of $V(\vec{K})$
we substitute $V^I(\vec{K})$, 
$V_0^S(\vec{K})$ and $V_M^S(\vec{K})$, respectively.
Taking into account (\ref{2.23a},b) we can separate the contributions
to $V_{\Lambda}^{out}$ into three parts,
\begin{eqnarray}
  V_{\Lambda}^{out}=V_{\Lambda,I}^{out}
  +V_{\Lambda,S_0}^{out}
  +V_{\Lambda,S_M}^{out} . 
  \label{2.27b}
\end{eqnarray}
Here $V_{\Lambda,I}^{out}$ is the contribution
from the interstitial region, while
$V_{\Lambda,S_0}^{out}$ and $V_{\Lambda,S_M}^{out}$
accounts for the contributions from the monopoles and the higher multipoles 
from the sites $\vec{n}' \neq \vec{n}$, respectively. 
Explicitly, we obtain 
\begin{mathletters}
\begin{eqnarray}
 & &  V_{\Lambda,I}^{out}=V_{\Lambda}^I ,
 \label{2.28a} \\
 & &  V_{\Lambda,S_0}^{out}= \left\{
 \begin{array}{l l} 
 V_0^{S_0}-\frac{q_0(R)}{R},  &  \Lambda = 0 ,  \\
 V_{\Lambda}^{S_0} , & \Lambda \neq 0, 
 \end{array} \right. 
 \label{2.28b} \\
 & &  V_{\Lambda,S_M}^{out} = \left\{
 \begin{array}{l l} 
   V_0^{S_M} , &   \Lambda = 0 ,  \\
   V_{\Lambda}^{S_M}-\frac{4\pi}{2l+1}
 \frac{q_{\Lambda}(R)}{R^{l+1}},  & \Lambda \neq 0 . 
 \end{array} \right. 
 \label{2.28c}
\end{eqnarray}
\end{mathletters}
The representation of $V(\vec{R})$ given by Eqs.\ (\ref{2.20}) and (\ref{2.27b}) 
is convenient because one clearly distinguishes the intra-site part 
$V_S$ from the inter- site ones,
$V_{\Lambda,S_0}^{out}$, $V_{\Lambda,S_M}^{out}$,
and $V_{\Lambda,I}^{out}$.   

\subsection {Comparison with the two-center expansion of the 
Coulomb potential}

It is instructive to compare the results of the previous subsection
with the two-center expansion of the Coulomb potential, \cite{Nik1,Nik2}
\begin{eqnarray}
 & & \frac{1}{|\vec{R}(\vec{n})-\vec{R}'(\vec{n}')|} \nonumber \\ 
 & & = \sum_{\Lambda \Lambda'} v_{\Lambda \Lambda'}(\vec{n},\vec{n}';\,r,r')\,
    S_{\Lambda}(\hat{r}(\vec{n}))\, S_{\Lambda'}(\hat{r}'(\vec{n}')) ,
 \label{2.30a} 
\end{eqnarray}
where
\begin{eqnarray}
 & & v_{\Lambda \Lambda'}(\vec{n},\vec{n}';\,r,r')\,
   =  \int \! d\Omega(\vec{n}) \int \! d\Omega(\vec{n}')\,
      {\frac{  S_{\Lambda}(\hat{n})\, S_{\Lambda'}(\hat{n}')}
     {|\vec{R}(\vec{n})-\vec{R}'(\vec{n}')|}} . \nonumber   \\
 & & \label{2.30b} 
\end{eqnarray}
The potential on the sphere surface at a site $\vec{n}$ is found
by integrating (\ref{2.30a}) over $\vec{r}'(\vec{n}')$ and summing over
the spheres $\vec{n}' (\neq \vec{n})$:
\begin{eqnarray}
 V_S^{out}(\Omega)=V_0^{out}+ {\sum_{\Lambda \neq 0}}'
 (V_{\Lambda,S_0}^{out} + V_{\Lambda,S_M}^{out} )\, S_{\Lambda}(\Omega) ,
 \label{2.30c} 
\end{eqnarray}
where (compare with Eqs.\ (\ref{2.28b},c))
\begin{mathletters}
\begin{eqnarray}
 & & V_{\Lambda,S_0}^{out}={\sum_{\vec{n}' \neq \vec{n}}}' 
  v_{\Lambda\,0}(\vec{n},\vec{n}',R) ,
\label{2.31a} \\
 & & V_{\Lambda,S_M}^{out}={\sum_{\vec{n}' \neq \vec{n}}}'
  {\sum_{\Lambda' \neq 0}}' 
  v_{\Lambda\,\Lambda'}(\vec{n},\vec{n}',R),
\label{2.31b} 
\end{eqnarray}
\end{mathletters}
and
\begin{eqnarray}
 v_{\Lambda\,\Lambda'}(\vec{n},\vec{n}',R)=
 \int_0^R v_{\Lambda \Lambda'}(\vec{n},\vec{n}';\,R,r') 
 \rho'_{\Lambda'}(r') r'^2 dr' .
 \label{2.32} 
\end{eqnarray}
Since \cite{Hei}
\begin{eqnarray}
 v_{\Lambda \Lambda'}(\vec{n},\vec{n}';\,r,r') 
 \sim \frac{(r)^l (r')^{l'}}{|\vec{X}(\vec{n})-\vec{X}(\vec{n}')|^{l+l'+1}},
 \label{2.33} 
\end{eqnarray}
the sums on $\vec{n}'$, Eqs.\ (\ref{2.31a},b), converge fast 
(except for the case $l=l'=0$ which is irrelevant here).
Practically, it is enough to consider
summations over a few neighboring shells.  

In many cases (for the cubic symmetry, for example)
the whole sum (\ref{2.31b}) is small in comparison with
(\ref{2.31a}). Eq.\ (\ref{2.31a}) then corresponds to the
potential calculated
with spherical symmetric charge distributions for the neighbors
($i.e.$ $\Lambda'=l'=0$)
and a homogeneous charge distribution in the interstices,
 Eqs. (\ref{2.3a}), (\ref{2.9a}). It follows from (\ref{2.33}) that
$v_{\Lambda 0}(\vec{n},\vec{n}';\,r,r')$ does not depend on $r'$
which is a dummy argument and 
\begin{eqnarray}
 v_{\Lambda\,0}(\vec{n},\vec{n}',R)=
 B_{\Lambda}(\vec{n}')\, Q_0(R) \frac{1}{\sqrt{4\pi}} .
 \label{2.34} 
\end{eqnarray}
Here $Q_0(R)$ is {\it the effective charge} 
inside the sphere given by Eq.\ (\ref{2.9a}),
and $B_{\Lambda}(\vec{n}')=v_{\Lambda 0}(\vec{n},\vec{n}';\,R,r')$.
Combining (\ref{2.31a}) and (\ref{2.34}) with (\ref{2.22}), 
we introduce the potential
\begin{mathletters}
\begin{eqnarray}
 V^{CEF}(\vec{r})=\frac{Q_0(R)}{\sqrt{4\pi}} 
 {\sum_{\Lambda \neq 0}}' 
  B_{\Lambda} \left( \frac{r}{R} \right)^l S_{\Lambda}(\hat{r}) ,
 \label{2.35a} 
\end{eqnarray}
where 
\begin{eqnarray}
  B_{\Lambda} = {\sum_{\vec{n}'}}' B_{\Lambda}(\vec{n}') .
 \label{2.35b} 
\end{eqnarray}
\end{mathletters}
The potential $V^{CEF}$ represents the crystal electric field 
potential (CEF) for
localized electrons at site $\vec{n}$
if there are no multipolar couplings with conduction electrons
and the electron density in the interstitial region is 
homogeneous, $V^I(\vec{K})=0$. 
Examples of calculations of $B_{\Lambda}$ for a face centered cubic
lattice (fcc) will be given in the next section.
In the nearest neighbor approximation one has 
$B_{\Lambda}=N_{nn} B_{\Lambda}(\vec{n}')$, where $\vec{n}'$ refers
to any of the nearest neighbors and 
$N_{nn}$ is the number of nearest neighbors.
Such CEF potential has been
considered in Refs.~\onlinecite{Nik1,Nik3}. 
A more sophisticated CEF for localized and itinerant electrons 
will be discussed in Sec.\ IV.

\section {Application to the face centered cubic lattice} 
\label{sec:pe} 

In this section we apply the technique of Sec.\ II to
test the method and to calculate some important structural constants.
As a Bravais lattice we take the face centered cubic (fcc) lattice.

\subsection{ Periodic Coulomb potential of
the monopole ($l=0$) charge density}

We first consider the potential of the unit ($q=1$) point
charges situated on a face centered cubic lattice.
For simplicity we take the cubic lattice constant $a=1$. (The volume
of the unit cell is $v=1/4$.)
We consider touching spheres on the fcc lattice with the 
close contact radius $R=\sqrt{2}/4$.
As we know from Sec.\ II, the crystal potential is defined only 
if the total charge of the unit cell is zero. 
Therefore, complementary to the positive point charges
we must introduce a compensating
negative charge density distributed homogeneously 
in the interstitial
region. 
The total charge confined in the interstitial region 
of the unit cell must be $q_I=-1$ with the density
$\rho_I=-1/v_I=-4/(1-\pi /3 \sqrt{2})$. 

Using the method of Sec.\ II
we first expand the potential of the interstitial region in Fourier
series and then extend it smoothly to the whole crystal, Eqs.\ (2.17a,b).
There are no high multipole moments inside a sphere and 
from Eq.\ (\ref{2.16a}) we find $V_M^S(\vec{K})=0$.
Due to electroneutrality the terms with $\vec{K}=0$ are omitted.
Since in the interstitial region there is only one component
$\rho_I(\vec{K}=0)$ which is different from zero, this results in 
$V^I(\vec{K} \neq 0)=0$.
Therefore, the potential in the interstices is given by (\ref{2.17a})
where the only nonzero contribution to the Fourier component $V(\vec{K})$
is the Ewald contribution $V_0^S(\vec{K})$, Eq.\ (\ref{2.15}).
Notice that for the Ewald expansion
we use an effective point charge $Q_0 \equiv Q_{eff}=q-\rho_I 4\pi R^3/3=
1+\pi/(3\sqrt{2}-\pi)$. The renormalization of charges is the price 
that we have to pay for
the extension of the Fourier expansion inside the spheres, Sec.\ II.

The potential inside a sphere $S(\vec{n})$ then is given by
\begin{eqnarray}
 V(\vec{r})=\frac{1}{r}+V_0^{out}+
 V_{04} \left(\frac{r}{R} \right)^4 K_4(\hat{r}) & + &
 V_{06} \left(\frac{r}{R} \right)^6 K_6(\hat{r})   \nonumber \\
 & & +...\; . \label{7.2} 
\end{eqnarray}
Here we introduce the cubic
harmonics $K_4(\Omega)$ and $K_6(\Omega)$ which are 
SAFs $S_{l=4,A1g}(\Omega)$ and 
$S_{l=6,A1g}(\Omega)$, correspondingly, \cite{Bra,rem} and
\begin{eqnarray}
 V_0^{out}=V_0-\frac{1}{R} .
  \label{7.1} 
\end{eqnarray}
It is interesting to remark that although we have started with
the model of point charges on the fcc lattice, 
the total crystal potential (\ref{7.2})
acquires contributions of the high multipole orders, $l=4,6,8,...$.
The constants $V_0$, $V_{04}$, $V_{06}$ of the potential 
are found from the boundary conditions on
the sphere surface, Eqs.\ (\ref{2.19a},b).
The potential energy of the charges (fcc lattice)
is given by
\begin{eqnarray}
  E_{fcc}/N=V_0^{out}+
  \frac{1}{2} \langle V _I \rangle =-3.9885 \; \mbox{ a.u.} , 
  \label{7.3} 
\end{eqnarray}
where $\langle V _I \rangle$ is the average potential of the
interstitial region $I$,
\begin{eqnarray}
  \langle V _I \rangle = \frac{1}{v_I} 
  \int_{I} V(\vec{R}) d\vec{R}=\frac{v}{v_I}
  {\sum_{\vec{K} \neq 0}}' V(\vec{K}) {\cal O}(\vec{K}) . 
  \label{7.4} 
\end{eqnarray}
The overlap integral ${\cal O}(\vec{K})$ reads
\begin{eqnarray}
 {\cal O}(\vec{K})=\delta_{\vec{K}}
 -\frac{4\pi R^2}{v} \frac{j_1(KR)}{K} .
  \label{7.4b} 
\end{eqnarray}
Thus $E_{fcc}/N$, Eq.\ (\ref{7.3}),  has the 
meaning of the Madelung constant for the fcc lattice. 

The values $V_0$, $V_{04}$, $V_{06}$ and $\langle V _I \rangle$ 
are proportional to the effective
charge $Q_{eff} \neq 1$. It is convenient to consider the 
normalized to the unit charge quantities
$\tilde{V}_0=V_0/Q_{eff}$, $\tilde{V}_{04}=V_{04}/Q_{eff}$,
$\tilde{V}_{06}=V_{06}/Q_{eff}$ and $\langle \tilde{V} _I \rangle=
\langle V_I \rangle/Q_{eff}$.
We study the convergence of 
$\tilde{V}_0$, $\tilde{V}_{04}$,
$\tilde{V}_{06}$ and $\langle \tilde{V} _I \rangle$ by employing
the Ewald method and the technique of Weinert,
Ref.~\onlinecite{Wei}. The results are quoted in Tables I and II
where the same number of reciprocal
vectors $|\vec{K}| < K_{max}$ is used for both approaches.
%
\begin{table} 
\caption{
 Comparison of the Ewald expansion and the method of Ref.~18.
 $G$ is the Ewald cut-off parameter, $n$ is the parameter of 
 Ref.~18.
 ``Exact" corresponds to the Ewald expansion with $K R \leq 1779$
 and $G=780$, with the accuracy of $\approx 10^{-5}$.
\label{tab1}     } 
 
 \begin{tabular}{c | c c c | c c c} 
  & \multicolumn{3}{c|}{Ewald} & \multicolumn{3}{c}{Ref.~\onlinecite{Wei}} \\
  $K_{max}R$ & $G$ & $\tilde{V}_0$ & $\langle \tilde{V} _I \rangle$ &
  $n$ & $\tilde{V}_0$ & $\langle \tilde{V} _I \rangle$ \\
\tableline
 11.54 &   7    & -0.4545 & -0.5607 & 5  & -0.4996 & -0.5607 \\
 14.74 &   8    & -0.5131 & -0.6200 & 8  & -0.5597 & -0.6200 \\
 28.45 &  12.5 & -0.6288 & -0.7359 & 17 & -0.6287 & -0.7359 \\
 46.33 &  20  & -0.6778 & -0.7849 & 39 & -0.6714 & -0.7849 \\
 68.54 &  30  & -0.6953 & -0.8024 & 62 & -0.6849 & -0.8024 \\
 90.73 &  40  & -0.7014 & -0.8085 & 84 & -0.6911 & -0.8085 \\
\tableline
 exact  & -  & -0.70922 & -0.81630 & - &  -0.70922 & -0.81630 
\end{tabular} 
\end{table} 
%
%
\begin{table} 
\caption{
 Comparison of the Ewald expansion and the method of Ref.~18.
 $G$ is the Ewald cut-off parameter, $n$ is the parameter of 
 Ref.~18.
\label{tab2}     } 
 
 \begin{tabular}{c | c c c | c c c} 
    & \multicolumn{3}{c|}{Ewald} & \multicolumn{3}{c}{Ref.~\onlinecite{Wei}} \\
  $K_{max}R$ & $G$ & $\tilde{V}_{04}$ & $\tilde{V}_{06}$ &
  $n$ & $\tilde{V}_{04}$ & $\tilde{V}_{06}$ \\
\tableline
 11.54 &   7    & -0.18135 & -0.14304 & 5  & -0.18233 & -0.14600 \\
 14.74 &   8    & -0.18186 & -0.14459 & 8  & -0.18192 & -0.14459 \\
 28.45 &  12.5 & -0.18193  &  -0.14467 & 17 & -0.18192 & -0.14466  \\
 46.33 &  20  &  -0.18192 & -0.14466 & 39 & -0.18192 & -0.14466 \\
\tableline
 exact  & -  & -0.18192 & -0.14466 & - &  -0.18192 & -0.14466
\end{tabular} 
\end{table}
(Though we did not optimize the convergence of Fourier series
by choosing the best $G$ and $n$ for the Ewald and Weinert
approaches.) 
We observe that $\tilde{V}_{06}$
and $\tilde{V}_{04}$ converge faster than  
$\tilde{V}_0$ and $\langle \tilde{V} _I \rangle$.

The Ewald method estimates better the monopole terms 
$\tilde{V}_0$ and $\langle \tilde{V} _I \rangle$, Table I.
In this case it is simple and stable, while
Weinert' procedure
requires evaluation of spherical Bessel functions of 
higher order for each $\vec{K}$.  

Both methods give almost the same results
for $\tilde{V}_{04}$ and $\tilde{V}_{06}$, Table~II.
$\tilde{V}_{04}$ and $\tilde{V}_{06}$ can be calculated
independently
by means of the two-center expansion of Coulomb interaction,
which we have considered in Sec.\ II D. 
In that case we first calculate 
$v_{04}(\vec{n},\vec{n}')$ and $v_{06}(\vec{n},\vec{n}')$,
Eq.\ (\ref{2.30b}), where $r=r'=R$, with
$S_0=1/\sqrt{4\pi}$ on a first site $\vec{n}$, and the cubic
harmonic (SAF) $K_4$ or $K_6$ on a second site $\vec{n}'$. 
Since the integrals 
$v_{04}(\vec{n},\vec{n}')$ and $v_{06}(\vec{n},\vec{n}')$
decrease very fast with the distance $d$ between the two sites 
($1/d^5$ for $K_4$ and $1/d^7$ for $K_6$), we obtain 
 $\tilde{V}_{04}$ and $\tilde{V}_{06}$
by summing $v_{04}(\vec{n},\vec{n}')$ and 
$v_{06}(\vec{n},\vec{n}')$ over all neighbors 
$\vec{n}'$ on the fcc lattice, {\it i.e.}
\begin{eqnarray}
  \tilde{V}_{0t}=\frac{1}{\sqrt{4\pi}}
  {\sum_{\vec{n}'} }' v_{0t}(\vec{n},\vec{n}') ,
  \label{new1} 
\end{eqnarray}
where $t=4$ or 6.
The results of such calculations are shown in Tables~III and IV.
%
\begin{table} 
\caption{
 Calculation of $\tilde{V}_{04}$ by summation over neighboring shells,
 $N_{sh}$ is the number of neighbors in the shell,
 $\vec{n}'$ stands for the coordinates of a shell representative,
 $v_{04}$ is the two-center integral (2.33) for
 $S_0$ and $K_4$. ``Exact" means summation over 128428 neighbors
 ($d\leq 12$) on the fcc lattice. 
\label{tab3}     } 
 
 \begin{tabular}{c r c r c} 
  shell & $N_{sh}$  & $\vec{n}'$  & $v_{04}$ & $\tilde{V}_{04}$  \\
\tableline
  1 & 12 & $(\frac{1}{2},\frac{1}{2},0)$ & -0.070694 & -0.23930 \\
  2 & 6  & $(1,0,0)$                     &  0.049988 & -0.15470 \\
  3 & 24 & $(\frac{1}{2},\frac{1}{2},1)$ & -0.004535 & -0.18540 \\
  4 & 12 & $(1,1,0)$                     & -0.002209 & -0.19288 \\
  5 & 24 & $(\frac{3}{2},\frac{1}{2},0)$ &  0.002782 & -0.17405 \\
\tableline
 10 &   &   &  & -0.18073 \\ 
 20 &   &   &  & -0.18283 \\
\tableline
 exact &   &  &  & -0.18191 
\end{tabular} 
\end{table} 
%
%
\begin{table} 
\caption{
 Calculation of $\tilde{V}_{06}$ by summation over neighboring shells,
 $N_{sh}$ is the number of neighbors in the shell,
 $\vec{n}'$ stands for the coordinates of a shell representative,
 $v_{06}$ is the two-center integral (2.33) for
 $S_0$ and $K_6$. ``Exact" means summation over 111956 neighbors
 ($d \leq 9$) on the fcc lattice. 
\label{tab4}     } 
 
 \begin{tabular}{c r c r c} 
  shell & $N_{sh}$  & $\vec{n}'$ & $v_{06}$ & $\tilde{V}_{06}$  \\
\tableline
  1 & 12 & $(\frac{1}{2},\frac{1}{2},0)$ & -0.044247  & -0.14978 \\
  2 & 6  & $(1,0,0)$                     & 0.002407  & -0.14571 \\
  3 & 24 & $(\frac{1}{2},\frac{1}{2},1)$ & 0.000299  & -0.14368 \\
  4 & 12 & $(1,1,0)$                     & -0.000346 & -0.14485 \\
  5 & 24 & $(\frac{3}{2},\frac{1}{2},0)$ &  0.000005 & -0.14482 \\
\tableline
 10 &   &   &  & -0.14467 \\ 
 20 &   &   &  & -0.14467 \\
\tableline
 exact &   &  &  & -0.14466 
\end{tabular} 
\end{table} 
%
The convergent values $\tilde{V}_{04}$ and $\tilde{V}_{06}$ perfectly
match those obtained by the techniques of Ewald and Weinert, Table~II.
However, the Ewald and Weinert' approaches require 
less computer time and are more
efficient than the two-center expansion. 
Finally, we would like to remark that the calculated in Table~II
exact values $\tilde{V}_{04}$ and $\tilde{V}_{06}$ can be used
for cubic crystal electric field, Eq.\ (\ref{2.35a},b).
For an fcc crystal with the cubic lattice constant $a \neq 1$ 
we have $B_4=\sqrt{4\pi}\, \tilde{V}_{04}/a$
and $B_6=\sqrt{4\pi}\, \tilde{V}_{06}/a$.

Although we have analyzed both methods at the close contact radius
$R=\sqrt{2}/4$ it is also possible to assess the validity of
the two methods by decreasing the sphere radius $r \leq R$.
Both methods are designed to describe the potential in the interstitial
region. They can not approximate the potential correctly in the whole space.
Therefore, both of them are expected to fail at a certain
small radius, but their departure from the exact solution
is an indication of their accuracy. 
The results for $\tilde{V}_{04}(r)$ and $\tilde{V}_{06}(r)$
($K_{max} R$=90.73, $G$=40, $n$=84) at $r < R$ are shown in Fig. 2 and 3.
%
\begin{figure} 
\vspace{-0.8cm}
\centerline{ 
\epsfig{file=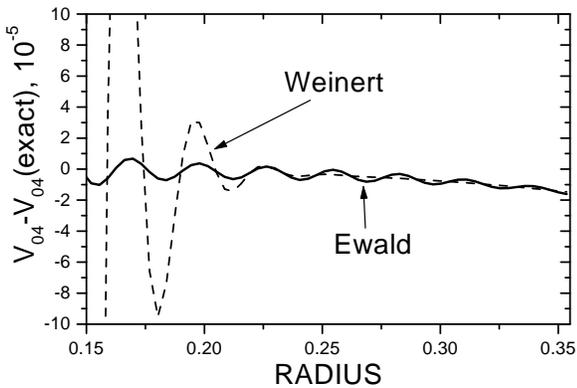,width=0.47\textwidth} 
} 
\vspace{-4.6cm}
\caption{
Comparison between the Ewald expansion and the method of 
Ref.~18 for $V_{04}(r)$.
} 
\label{fig2} 
\end{figure} 
%

%
\begin{figure} 
\vspace{-0.4cm}
\centerline{ 
\epsfig{file=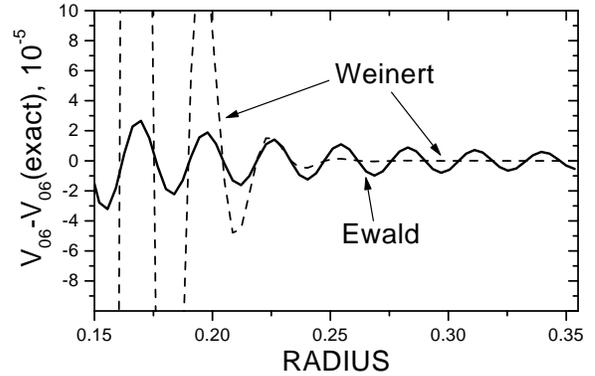,width=0.47\textwidth} 
} 
\vspace{-4.6cm}
\caption{
Comparison between the Ewald expansion and the method of 
Ref.~18 for $V_{06}(r)$.
} 
\label{fig3} 
\end{figure} 
Weinert' procedure gives more smooth values for $r$ close to $R$,
but the Ewald expansion is more stable in the whole range.
The calculation of the monopole term $V_0^{out}$ (or $V_0$) 
is of special interest.
By decreasing the sphere radius $r$ we change the density of
the compensating negative charge distribution. As a result, both
$V_0(r)$ and $V_0^{out}(r)$, Eq.\ (\ref{7.1}), are functions of $r$.
From the last value however
it is possible to reconstruct $V_0^{out}(R)$
at the close contact radius $R$. On can easily obtain the
result from electrostatic considerations:
\begin{mathletters}
\begin{eqnarray}
   \left. V_0^{out}(R) \right|_r=\frac{V_0^{out}(r)+2\pi(R^2-r^2)\rho}
   {1-\frac{4}{3} \pi(R^3-r^3)\rho} ,
  \label{7.5a} 
\end{eqnarray}
where 
\begin{eqnarray}
 \rho=\left( \frac{1}{4}-\frac{4\pi}{3}r^3 \right)^{-1} ,
  \label{7.5b} 
\end{eqnarray}
\end{mathletters}
In principle, 
$V_0^{out}(R)$ should be independent of $r$, but in practice
it exhibits such dependence as indicated on the left hand side of (\ref{7.5a}).
This is because the estimation (\ref{7.5a}) depends on the accuracy of 
the calculated value $V_0^{out}(r)$ and as such, it is a function of $r$.
We find it convenient to study $V_0^{out}(R)|_r$ which is expected
to be constant rather than the initial value $V_0^{out}(r)$.
In order to compare the two techniques we plot the function 
$\triangle(r)=(V_0^{out}|_r-V_0^{out}|_{ex})/V_0^{out}|_{ex}$
in Fig. 4. ($V_0^{out}|_{ex}=-5.5612$ is the best value for $V_0^{out}$
calculated from data of Table~I, last row.) 
As one clearly sees from Fig.~4,  
the Ewald expansion performs better than the approach of Weinert.
%
\begin{figure} 
\vspace{-1.4cm}
\centerline{ 
\epsfig{file=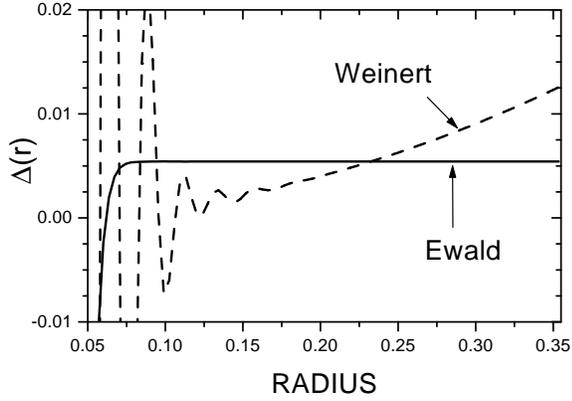,width=0.46\textwidth} 
} 
\vspace{-4.6cm}
\caption{
Comparison between the Ewald expansion ($G=40$, $KR<90.73$) and the method of 
Ref.~18 for $V_0(r)$ ($n=84$), see text for details.
} 
\label{fig4} 
\end{figure} 

\subsection{Periodic Coulomb potentials of 
the $l=4$ and $l=6$ cubic charge density}

Here we illustrate how the proposed method works for
high order multipoles with $l=4$ and $l=6$.
We consider two cases.
In the first case the charge distribution on the 
touching sphere surfaces is given by
\begin{mathletters}
\begin{eqnarray}
 \rho_4(\vec{r})=\frac{\delta(r-R)}{r^2} \, K_4(\hat{r}) ,
  \label{7.6a} 
\end{eqnarray}
In the second case it is
\begin{eqnarray}
 \rho_6(\vec{r})=\frac{\delta(r-R)}{r^2} \, K_6(\hat{r}) .
  \label{7.6b}  
\end{eqnarray}
\end{mathletters}
Here $\delta(r')$ is the one-dimensional delta-function and
$K_4$, $K_6$ are cubic harmonics (SAFs). \cite{rem}
Now it is not necessary to introduce a compensating charge
in the interstitial
region since the total charge of the unit cell given by integration of
(\ref{7.6a}) or (\ref{7.6b}) over the polar angles yields zero.
According to the treatment of Sec.\ II we start with the potential
of a single sphere, say $S(\vec{n}=0)$. 
The potentials then are
\begin{mathletters}
\begin{eqnarray}
  U_S^4(\vec{r})=\frac{4\pi}{9} \frac{1}{R} \left( \frac{r}{R} \right)^4 
 K_4(\hat{r}) , 
   \label{7.7a} \\ 
  U_S^6(\vec{r})=\frac{4\pi}{13} \frac{1}{R} \left( \frac{r}{R} \right)^6 
  K_6(\hat{r}) ,
  \label{7.7b}  
\end{eqnarray}
\end{mathletters}
for $r \leq R$, and 
\begin{mathletters}
\begin{eqnarray}
  U_S^4(\vec{r})=\frac{4\pi}{9} \frac{1}{R} \left( \frac{R}{r} \right)^5 
 K_4(\hat{r}) , 
   \label{7.8a} \\ 
  U_S^6(\vec{r})=\frac{4\pi}{13} \frac{1}{R} \left( \frac{R}{r} \right)^7 
 K_6(\hat{r}) ,
  \label{7.8b}  
\end{eqnarray}
\end{mathletters}
for $r \geq R$. Notice that the potentials $U_S^4$ and 
$U_S^6$ have discontinuous derivatives at $R$. 
(Here we do not make any distinction between $U(r)$ and ${\cal U}(r)$, Sec.\ II,
since there is no renormalization of the charge density
inside a sphere.)
Using the method of Sec.\ II we find from Eqs.\ (\ref{2.16a},b) the Fourier 
transforms of the potentials of the periodic arrangements 
of such charged surfaces on the fcc lattice,
\begin{mathletters}
\begin{eqnarray}
 & & V_4(\vec{K})=\frac{(4\pi)^2}{9\, v} \frac{R}{K}
 (j_3(KR)+j_5(KR))\,K_4(\vec{K}) , 
   \label{7.9a} \\ 
 & & V_6(\vec{K})=-\frac{(4\pi)^2}{13\, v} \frac{R}{K}
  (j_5(KR)+j_7(KR))\,K_6(\vec{K}) .
  \label{7.9b}  
\end{eqnarray}
\end{mathletters}
Notice, that we arrive at the same results if first we 
Fourier transform the densities,
\begin{mathletters}
\begin{eqnarray}
 & & \rho_4(\vec{K})=\frac{4\pi}{v} j_4(KR) \, K_4(\vec{K}) ,
  \label{7.10a} \\ 
 & & \rho_6(\vec{K})=-\frac{4\pi}{v} j_6(KR) \, K_6(\vec{K}) ,
  \label{7.10b}  
\end{eqnarray}
\end{mathletters}
and then calculate the potentials by  means of
\begin{mathletters}
\begin{eqnarray}
 V_4(\vec{K})=\frac{4\pi}{K^2} \rho_4(\vec{K}) , 
   \label{7.11a} \\ 
 V_6(\vec{K})=\frac{4\pi}{K^2} \rho_6(\vec{K})  .
  \label{7.11b}  
\end{eqnarray}
\end{mathletters}
(In oder to establish the equality with (\ref{7.9a},b) we used
the property 10.1.21 of Bessel functions, Ref.~\onlinecite{Abr}.)
As we have many times mentioned before (Sec.\ II) the Fourier
expansion is valid only in the interstitial region.

Inside a sphere $S(\vec{n})$ the first potential is given by 
\begin{eqnarray}
 & & 
 V_4(\vec{r})=V_{40}+V_{44} \left(\frac{r}{R} \right)^4 K_4(\hat{r}) 
 + V_{46} \left(\frac{r}{R} \right)^6 K_6(\hat{r}) +... .
 \nonumber \\
 & & \label{7.12}  
\end{eqnarray}
For the second potential, accordingly, we have
\begin{eqnarray}
 & & 
 V_6(\vec{r})=V_{60}+V_{64} \left(\frac{r}{R} \right)^4 K_4(\hat{r}) 
 + V_{66} \left(\frac{r}{R} \right)^6 K_6(\hat{r}) +... .
 \nonumber  \\
 & & \label{7.12'}  
\end{eqnarray}
The constants $V_{40}$, $V_{60}$, and $V_{44}$, $V_{46}$, $V_{64}$, $V_{66}$
are found from the boundary conditions, Eqs.\ (\ref{2.19a}) and (\ref{2.19b}),
where $V(\vec{K})$ is given by $V_4(\vec{K})$, Eq.\ (\ref{7.9a}),
for the first potential and by $V_6(\vec{K})$, Eq.\ (\ref{7.9b}), for
the second potential.
(Compare these results with the potentials (\ref{7.7a}) and (\ref{7.7b})
for a single sphere.)

In order to distinguish the potential due to the 
multipole moment of the sphere $S(\vec{n})$
from the potential of the rest of the crystal,
we subtract from $V_4$ and $V_6$
the contribution induced by the multipole moment of the sphere 
(see Eqs.\ (\ref{2.28a}-c)) and introduce the parameters
\begin{mathletters}
\begin{eqnarray}
 V_{44}^{out}=V_{44}- \frac{4\pi}{9}\frac{1}{R}  , 
   \label{7.12a} \\ 
 V_{66}^{out}=V_{66}- \frac{4\pi}{13}\frac{1}{R}   .
  \label{7.12b}  
\end{eqnarray}
\end{mathletters}

We have calculated $V_{44}^{out}$, $V_{66}^{out}$,
$V_{46}$ and $V_{60}$ and then compared them with those
obtained by the method of Ref.~\onlinecite{Wei}, see Tables~V and VI.
%
%
\begin{table} 
\caption{
 Comparison of the 
 direct Fourier expansion and the method of Ref.~18.
 $n_4$ and $n_6$ are the parameters of 
 Ref.~18 for $V_4$ and $V_6$.
 ``Exact" corresponds to Weinert' expansion with $K_{max}R \leq 124.08$
 and $n_4=96$, $n_6=94$.
\label{tab5}     } 
 
 \begin{tabular}{c | c c | c c c c} 
    & \multicolumn{2}{c|}{direct expansion} & \multicolumn{4}{c}{Ref.~\onlinecite{Wei}} \\
  $K_{max}R$ & $V_{44}^{out}$ & $V_{66}^{out}$ &
  $n_4$ & $V_{44}^{out}$ & $n_6$ & $V_{66}^{out}$  \\
\tableline
 35.19 & 1.8876 & 1.1750 & 24  & 2.2139 & 22 & 1.5161 \\
 57.37 & 2.0113 & 1.3074 & 46  & 2.2139 & 44 & 1.5334 \\
 79.60 & 2.0688 & 1.3696 & 69 & 2.2140 & 67 &  1.5220 \\
 101.8 & 2.1007 & 1.4036 & 91 & 2.2140 & 89 & 1.5211 \\
\tableline
 exact &  2.21402 & 1.52112 & $-$ &  2.21402 & $-$ & 1.52112 
\end{tabular} 
\end{table} 
%
%
\begin{table} 
\caption{
 Comparison of the 
 direct Fourier expansion and the method of Ref.~18.
 $n_4$ and $n_6$ are the parameters of 
 Ref.~18 for $V_4$ and $V_6$.
 ``Exact" corresponds to Weinert' expansion with $K_{max}R \leq 124.08$
 and $n_4=96$, $n_6=94$.
\label{tab6}     } 
 
 \begin{tabular}{c | c c | c c c c} 
  & \multicolumn{2}{c|}{direct expansion} & \multicolumn{4}{c}{Ref.~\onlinecite{Wei}} \\
  $K_{max}R$ & $V_{46} \cdot 10^2$ & $V_{60}$ &
  $n_4$ & $V_{46}\cdot 10^2$ & $n_6$ & $V_{60}$  \\
\tableline
 35.19 & -2.5778 & -0.1426 & 24  & -1.6667 & 22 & -0.1442 \\
 57.37 & -1.9364 & -0.1431 & 46  & -1.7042 & 44 & -0.1464 \\
 79.60 & -1.7908 & -0.1436 & 69 & -1.6820 & 67 &  -0.1448 \\
 101.8 & -1.7529 & -0.1439 & 91 & -1.6812 & 89 & -0.14467 \\
\tableline
 exact &  -1.68119 & -0.14466 & $-$ &  -1.68119 & $-$ & -0.14466 
\end{tabular} 
\end{table} 
Alternatively, we can compute $V_{44}^{out}$
and $V_{66}^{out}$ by performing the two-center multipole expansions 
(\ref{2.30a}), (\ref{2.30b}). To do it, 
we first calculate integrals $v_{44}(\vec{n},\vec{n}')$ and $v_{66}(\vec{n},\vec{n}')$,
Eq.\ (\ref{2.30b}), between the functions $K_4(\vec{n})$ and $K_4(\vec{n}')$, 
or between $K_6(\vec{n})$ and $K_6(\vec{n}')$ at sites $\vec{n}$ and $\vec{n}'$. 
By summing over ${\vec{n}'}$ on the fcc lattice we estimate $V_{44}^{out}$
and $V_{66}^{out}$, {\it i.e.}
\begin{eqnarray}
  V_{tt}^{out}= 
  {\sum_{\vec{n}'} }' v_{tt}(\vec{n},\vec{n}') ,
  \label{new2} 
\end{eqnarray}
where $t=4$ or 6.
Clearly, the accuracy depends on the number
of sites $\vec{n}'$ included in the summations.
The results of these calculations are given in Table VII. 
(The representative coordinates of five nearest shells as well as 
the number of atoms of the shells are quoted in Tab.\ III.)
%
%
\begin{table} 
\caption{
 Calculation of $V_{44}$ and $V_{66}$ by summation over neighboring shells.
 The discrepancy between the converged value for $V_{66}$ and the exact one
 from Table VI is due to numerical errors of the two center integration.
 \label{tab7}     } 
 
 \begin{tabular}{c | r c | r c} 
  shell & $v_{44}$  & $V_{44}^{out}$ & $v_{66}$ & $V_{66}^{out}$  \\
\tableline
  1 & 0.17897  & 2.14767 & 0.12632  & 1.51593 \\
  2 & 0.01406  & 2.23204 & 0.00064  & 1.51976 \\
  3 & -0.00091 & 2.21024 & 0.00005  & 1.52088 \\
  4 & 0.00035  & 2.21444 & 0.00002 & 1.52106 \\
  5 & -0.00002 & 2.21393 & 0.0 & 1.52102 \\
\tableline
 10 & $-$  & 2.21403  & $-$ & 1.52102 \\ 
 20 & $-$  & 2.21401  & $-$ & 1.52102 \\
\tableline
 exact & $-$ & 2.21402  &   $-$  & 1.52112 
\end{tabular} 
\end{table}

From the Tables V-VII it follows that the method of Ref.\ \onlinecite{Wei}
gives a better convergence {\it despite the fact that the expressions}
(\ref{7.10a},b), (\ref{7.11a},b) {\it for the Fourier
coefficients are exact}. Interestingly,
the exact expressions formally
correspond to $n_4=n_6=-1$ in Eq.\ 28 of Ref.\ \onlinecite{Wei},
while the better convergence is achieved for large positive $n_4$ and $n_6$
implied by the technique of Ref.~\onlinecite{Wei}.
We ascribe this property to the fact that the potentials $V_4$
and $V_6$ have discontinues derivatives at the sphere boundaries.
Therefore, the exact expansions  (\ref{7.10a},b), (\ref{7.11a},b) try
to reproduce these cusps while
the pseudo-charge density of Ref.~\onlinecite{Wei}  
has $n-1$ continuous derivatives thereby avoiding the problem.
Thus, our test calculations for the densities 
(\ref{7.6a}) and (\ref{7.6b}) indicate that
Weinert' procedure gives a better convergence. 

We can also generalize the present consideration for 
arbitrary multipole sphere moments $Q_4(R)$ and $Q_6(R)$ 
given by Eq.\ (\ref{2.11a}) on an fcc lattice when 
its cubic lattice constant $a \neq 1$.
For the former case we obtain that the potential inside a sphere due
to the other spheres' multipoles $Q_4$ of the fcc crystal is
\begin{eqnarray}
 & & 
 V_4^{out}(\vec{r})=B_{40}+B_{44} \left(\frac{r}{R} \right)^4 K_4(\hat{r}) 
 + B_{46} \left(\frac{r}{R} \right)^6 K_6(\hat{r}) +... ,
 \nonumber \\
 & & \label{7.14}  
\end{eqnarray}
where $B_{40}=k_4 V_{40}$, $B_{44}=k_4 V_{44}^{out}$, $B_{46}=k_4 V_{46}$, and
the coefficient of proportionality for touching spheres ($R=\sqrt{2}a/4$) 
is given by
\begin{eqnarray}
 k_4=\frac{64}{a^5} Q_4(R) .
  \label{7.15}
\end{eqnarray}
Analogously, the expression for the potential 
$V_6^{out}$ due to $Q_6(R)$ is obtained as
\begin{eqnarray}
 & & 
 V_6^{out}(\vec{r})=B_{60}+B_{64} \left(\frac{r}{R} \right)^4 K_4(\hat{r}) 
 + B_{66} \left(\frac{r}{R} \right)^6 K_6(\hat{r}) +... ,
 \nonumber \\
 & & \label{7.16}  
\end{eqnarray}
where $B_{60}=k_6 V_{40}$, $B_{64}=k_6 V_{64}$, $B_{66}=k_6 V_{66}^{out}$, and
\begin{eqnarray}
 k_6=\frac{512}{a^7} Q_6(R) .
  \label{7.17}
\end{eqnarray}
Notice that here we can immediately determine the potentials using exact
values $V_{40}=\tilde{V}_{04}$, $V_{44}$, $V_{46}$,
$V_{60}=\tilde{V}_{06}$, $V_{64}$ and $V_{66}$ quoted in Tables~V and VI.
These parameters as well as those calculated in Sec.\ III.A
are in fact important structural constants of the fcc lattice.

\section {Hartree-Fock-Roothaan method:
          Calculations of the direct matrix elements }
 
\label{sec:HFR}

The Hartree-Fock operator for the electronic system 
is defined as \cite{Roo,HFR}
\begin{eqnarray}
   {\cal F}=\sum_d \left( {\cal H}_d + {\cal J}_d - {\cal K}_d \right) ,
 \label{3.1}
\end{eqnarray}
where ${\cal H}_d$ is the Hamiltonian operator for a $d$th
electron moving in the field of nuclei alone, 
\begin{eqnarray}
   {\cal H}_d=-\frac{\vec{\nabla}^2_d}{2} 
   -\sum_{\vec{n}} \frac{Z}{|\vec{R}_d-\vec{X}(\vec{n})|} ,
 \label{3.2}
\end{eqnarray}
${\cal J}_d$ and ${\cal K}_d$ are
the direct Coulomb and exchange operators, respectively.
In order to obtain the best trial function 
of the $d$th electron, $\psi_d(x)=\langle x | d \rangle$,
one considers basis functions $\chi_a(x)=\langle x | a \rangle$,
and expands $\psi_d$ in terms of $\chi_a$,
\begin{eqnarray}
  \psi_d(x) =\sum \gamma_{da}\, \chi_a(x) ,
  \label{3.3}
\end{eqnarray}
where $\gamma_{da}$ are the coefficients of the expansion, and the index $x$
stands for the coordinates $\vec{R}$ and the spin projection $s_z=\pm1/2$
of the electron, $x \equiv (\vec{R},s_z)$. 
For the matrix elements of ${\cal J}_d$ and 
${\cal K}_d$ one has
\begin{mathletters}
\begin{eqnarray}
 & &  \langle a | {\cal J}_d | b \rangle = 
   {\sum_c}' \int \chi_a^*(x_1) \chi_b(x_1) \frac{1}{R_{12}}
   \psi_c^*(x_2) \psi_c(x_2) dx_1 dx_2 , \nonumber \\
 & & \label{3.4a} \\
 & &  \langle a | {\cal K}_d | b \rangle = 
   {\sum_c}' \int \chi_a^*(x_1) \psi_c(x_1) \frac{1}{R_{12}}
   \chi_b(x_2) \psi_c^*(x_2) dx_1 dx_2 . \nonumber \\
 & & \label{3.4b}
\end{eqnarray}
\end{mathletters}
In Eqs.\ (\ref{3.4a},b) the summation is understood
over all occupied electron states except $d$.
In the following we consider
the restricted HF method. 
Usually in the HF method \cite{Roo} the wave function
$\psi_d(x)=\phi(\vec{R})\zeta(s_z)$ is given by the coordinate 
(orbital) part $\phi(\vec{R})$
and the spinor part $\zeta(s_z)$. 
However, it is well known that
the spin-orbit coupling mixes the two components 
$\zeta_+=\zeta(1/2)$ and $\zeta_-=\zeta(-1/2)$.
(The spin-orbit coupling is especially important for core shells.)
Therefore, here we consider more general spin-orbitals,
\begin{eqnarray}
 \psi_d(x)=\phi_+(\vec{R})\,\zeta_+ + \phi_-(\vec{R}) \zeta_- .    
 \label{3.5a}
\end{eqnarray}
Using the time reversal symmetry ${\cal T}$, \cite{Tin} one can construct from
$\psi_d$ the time reversed state $\psi^*_d={\cal T}\psi_d$
(see Appendix C).
According to the Kramers theorem $\psi_d$ and $\psi^*_d$ have
the same energy $\epsilon_d$ and are orthogonal. 
Therefore, such treatment is
analogous to the conventional restricted HF method \cite{Roo}
with the double occupancy of $\epsilon_d$. 
However, in the following we will treat these two states
separately, as different components of a 
doubled valued irreducible representation. Therefore,
although we work within the restricted HF method,
there are no factor 2 in front of ${\cal H}_d$ and ${\cal J}_d$
in Eq.\ (\ref{3.1}).
Notice, that as a consequence of (\ref{3.5a}),
the integrals (\ref{3.4a},b) include summations over two
spin components.

We regroup ${\cal F}$ as
\begin{eqnarray}
   {\cal F}={\cal F}^{Coul}-{\cal F}^{exc} ,
 \label{3.5}
\end{eqnarray}
where
\begin{eqnarray}
   {\cal F}^{Coul}=\sum_d ( {\cal H}_d + {\cal J}_d ) ,
 \label{3.6}
\end{eqnarray}
and
\begin{eqnarray}
   {\cal F}^{exc}=\sum_d {\cal K}_d .
 \label{3.7}
\end{eqnarray}
The matrix elements of ${\cal F}^{Coul}$ can be written in the following form:
\begin{eqnarray}
  \langle a | {\cal F}^{Coul} | b \rangle =
  \int \chi_a^*(x) 
  \left(-\frac{1}{2}\vec{\nabla}^2+ V^d(\vec{R}) \right) 
  \chi_b(x) \, dx ,
 \label{3.8}
\end{eqnarray}
where the Hartree (electrostatic) potential $V^d$ 
acting on the $d$th electron reads
\begin{eqnarray}
 & & V^d(\vec{R})  = {\sum_c}'
 \int \frac{\psi_c^*(x') \psi_c(x')}{|\vec{R}-\vec{R}'|} dx' 
 -\sum_{\vec{n}} \frac{Z}{|\vec{R}-\vec{X}(\vec{n})|} . \nonumber \\
 & & \label{3.9}
\end{eqnarray}
It is convenient to rewrite the potential as
\begin{eqnarray}
 V^d(\vec{R})  = V(\vec{R})-{\cal V'}^d(\vec{R}) ,
 \label{3.10}
\end{eqnarray}
where $V(\vec{R})$ is the potential of all electrons and nuclei
considered in section II,
\begin{eqnarray}
 V(\vec{R})=
 \int \frac{\rho(\vec{R}')}{|\vec{R}-\vec{R}'|} d\vec{R}' ,
 \label{3.11}
\end{eqnarray}
while ${\cal V'}^d$ is the potential created by the $d$th electron alone,
\begin{eqnarray}
 {\cal V'}^d(\vec{R})=
 \int \frac{\psi_d^*(x') \psi_d(x')}{|\vec{R}-\vec{R}'|} dx' .
 \label{3.12}
\end{eqnarray}
The charge density $\rho(\vec{R})$ in (\ref{3.11}),
\begin{mathletters}
\begin{eqnarray}
 \rho(\vec{R})=\rho_{el}(\vec{R})+
 \rho_n(\vec{R}) ,  
 \label{3.12b}
\end{eqnarray}
comprises the point charges of nuclei,
\begin{eqnarray}
 \rho_n(\vec{R})= \sum_{\vec{n}} Z \delta(\vec{R}-\vec{X}(\vec{n})) ,  
 \label{3.12c}
\end{eqnarray}
and the total electronic charge density,
\begin{eqnarray}
 \rho_{el}(\vec{R})= {\sum_c} \psi_c^*(x) \psi_c(x) .
 \label{3.12'}
\end{eqnarray}
(Again, in (\ref{3.12'}) the summation is implied over two spin components.)
In solids we are dealing with the two types of electron states, which are
extended states of valence electrons 
(with the total charge density $\rho_{val}(\vec{R})$) and localized
states of core electrons ($\rho_{core}(\vec{r}(\vec{n}))$), and
\begin{eqnarray}
  \rho_{el}(\vec{R})= \rho_{val}(\vec{R}) + 
  \sum_{\vec{n}} \rho_{core}(\vec{R}-\vec{X}(\vec{n})) .
 \label{3.14} 
\end{eqnarray}
\end{mathletters}
The Hartree-Fock-Roothaan method implies a self-consistent field
procedure. \cite{Roo,HFR} In particular, for itinerant states
($d=(\vec{k},\alpha)$) one solves the secular equation
\begin{eqnarray}
  & & \sum_a {\cal F}_{ba}(\vec{k},\alpha)\, \gamma_a(\vec{k},\alpha)
   =E(\vec{k},\alpha) \sum_a {\cal O}_{ba}(\vec{k})\, 
   \gamma_a(\vec{k},\alpha) . \nonumber \\
 & & \label{3.14a} 
\end{eqnarray}
Here $\vec{k}$ is the wave vector, $\alpha$ is the band index,
${\cal O}$ is the overlap matrix and $E(\vec{k},\alpha)$ is the
corresponding HF energy. \cite{Roo}
The HF operator ${\cal F}$ depends implicitly on
the solution (via the coefficients $\gamma_a$). Below in this section
we calculate the matrix elements of ${\cal F}^{Coul}$, while in
Sec.\ V we consider the matrix elements of ${\cal F}^{exc}$.

\subsection{Multipole expansion of electron density}

In order to calculate the matrix elements of 
${\cal F}^{Coul}$ for a $d$th electron
from Eqs.\ (\ref{3.8}) we have to know the electrostatic
potential $V^d$, Eq.\ (\ref{3.9}) or (\ref{3.10}).
We have already thoroughly studied this problem in Sec.\ II and
know how to proceed
starting with the dual representation (\ref{2.2a},b) of charge density.
The only problem is to construct the charge density
of the itinerant and the localized electrons.
In this subsection we do it explicitly for the core (A1) and
the valence (A2) electrons.

{\bf A1}. 
The density of the inner closed shells at site ${\vec n}$ reads
\begin{mathletters}
\begin{eqnarray}
  \rho_{core}(\vec{r}(\vec{n}))=\sum_{\tau} n_{\tau}
   \rho_{\tau}(\vec{r}(\vec{n})) ,
 \label{3.23} 
\end{eqnarray}
where
\begin{eqnarray}
  \rho_{\tau}(\vec{r})= 
 | \langle \vec{r},s_z | \tau \rangle |^2 =
 |\langle x | \tau \rangle|^2 .
 \label{3.23b} 
\end{eqnarray}
\end{mathletters}
Here $\langle x | \tau \rangle$ stands for the wave functions of
localized electrons, and  
$n_{\tau}=1$ if $\epsilon(\tau)<E_F$ and zero
otherwise ($E_F$ is the Fermi energy). 
In (\ref{3.23}) summation over $s_z=\pm1/2$ is implied.
Here we will not consider the case of partially filled
core shells. (Such situation occurs in lanthanides
with localized 4$f$ electrons or in transition elements with $d$ electrons.)
Furthermore,
we assume that all core electrons at site $\vec{n}$ are confined 
inside the sphere $S(\vec{n})$ so that the core wave functions
with their first derivatives fall to zero at $R$.

The inner closed shells are classified according to the principal
quantum number $n$, the total angular
momentum (including spin) $j$ and the orbital number $l$.
In the case of the spherical symmetry these $(2j+1)$ electronic states
belong to a doubled valued irreducible representation $D_j$ of 
the 3-dimensional rotation group $SO(3)$. \cite{Bra}
In the presence of crystal environment these levels (except $s$)
in general are split into doubled valued 
irreducible representations $(\Gamma,\nu)$
of a crystal double group, \cite{Tin}
\begin{eqnarray}
  D_j \rightarrow \sum \Gamma(\nu) ,
 \label{3.24} 
\end{eqnarray}
where $\nu$ is used to label representations which occur more than once.
We shall classify these core electronic states according to the atomic indices
$n$, $l$, $j$, and $\Gamma$, $\nu$, $k$ ($k$ labels the rows of $\Gamma$) 
{\it i.e.} $\tau \equiv (n,l,j;\, \Gamma,\nu,k)$, and 
\begin{eqnarray}
   \langle x | \tau \rangle
   = {\cal R}_{\tau}(r) \langle \hat{r},s_z | \tau \rangle=
  {\cal R}_{\tau}(r) \langle \hat{r},s_z  | n,l,j;\, \Gamma, \nu, k \rangle .
 \label{3.25} 
\end{eqnarray}
The orientational (spin-orbital) functions are linear combinations
of real SAFs and spinors,
\begin{eqnarray}
 \langle \hat{r} | n,l,j;\, \Gamma, \nu, k \rangle 
 = \sum c_{\tau}(\lambda,s_z) S_{\lambda}(\Omega) \zeta(s_z) . 
 \label{3.25b} 
\end{eqnarray}
In the case of the spherical symmetry $\lambda=(l,m)$,
$\tau=(n,j,m_j)$ and $c_{\tau}(\lambda,s_z)$
are Clebsch-Gordan (or Wigner) coefficients \cite{Tin,Bra}
but for other point groups these coefficients are not well known.
We have derived $c_{\tau}(\lambda,s_z)$ for
the cubic double group $O'_h$ in Appendix C.
Functions
$\langle \hat{x} | \tau \rangle \equiv \langle \hat{r},s_z | \tau \rangle$ 
are independent of the
radial part ${\cal R}_{n,j,l}(r)$, which is assumed to be the same
for all states belonging to the same $n$, $j$ and $l$.
The electronic states $| n,l,j;\, \Gamma, \nu, k \rangle$ distinguished by
the index $k$ have the same energy, $\epsilon(l,j;\Gamma,\nu)=\epsilon(\tau)$.
We next consider the multipole expansion of $\rho_{\tau}(\vec{r})$, 
Eq.\ (\ref{3.23b}), and find
\begin{eqnarray}
  \rho_{\tau}(\vec{r}) 
  = \sum_{\Lambda} {\cal R}_{\tau}^2(r) c_{\Lambda}(\tau,\tau) 
  S_{\Lambda}(\hat{r}) .
 \label{3.26} 
\end{eqnarray}
Since the core density is invariant of the point group, 
only the full symmetrical irreducible representations
has to be considered, with $\Gamma_{\Lambda}=A_{1g}$. 
The coefficients $c_{\Lambda}(\tau,\tau')$ are given by
\begin{eqnarray}
 c_{\Lambda}(\tau,\tau') = \sum_{s_z=-1/2}^{+1/2} 
 \int \langle \Omega,s_z | \tau \rangle^*  S_{\Lambda}(\Omega) 
  \langle \Omega,s_z | \tau' \rangle d\Omega .
 \label{3.27} 
\end{eqnarray}
(We recall that $\Omega$ or $\hat{r}$ stand for the two polar
angles $(\Theta,\phi)$.)

{\bf A2}. 
The density of the valence electrons reads
\begin{eqnarray}
  \rho_{val}(\vec{R})=\sum_{\vec{k},\alpha} n_{\vec{k} \alpha}
 | \langle x | \vec{k} \alpha \rangle |^2 ,
 \label{3.15} 
\end{eqnarray}
where $\langle x | \vec{k} \alpha \rangle$ is the
wave function of a delocalized electron with
the wave vector $\vec{k}$ and the band index $\alpha$;
$n_{\vec{k} \alpha}$ is the occupation number, 

We expand the electron density of itinerant electrons in
 multipole series inside the spheres.
In the following we will use the LAPW basis functions, Appendix B. 
Using Eqs.\ (\ref{b.7}) we find that the
local density at a site $\vec{n}$ is given by
\begin{eqnarray}
 \rho_{val}(\vec{r}(\vec{n}))=
 \sum_{\Lambda} \rho_{\Lambda}(r(\vec{n})) S_{\Lambda}(\hat{r}(\vec{n})) ,
 \label{3.16}
\end{eqnarray}
where
\begin{eqnarray}
 \rho_{\Lambda}(r)=\sum_{l_1 l_2} 
 u_{l_1}^p(r) u_{l_2}^t(r) 
 \rho_{\Lambda}^{(l_1,p) (l_2,t)}   .
 \label{3.17}
\end{eqnarray}
Here and below summation is understood over the repeated indices
$p$ and $t$, and
\begin{eqnarray}
  \rho_{\Lambda}^{(l_1,p) (l_2,t)}  =
 \frac{1}{N} \sum_{\vec{k},\alpha} n_{\vec{k}, \alpha}
 c_{\Lambda}^{(l_1,p) (l_2,t)}(\vec{k},\alpha; \vec{k},\alpha) .
 \label{3.18}
\end{eqnarray}
(See Appendix B for definitions.)
The coefficients $c_{\Lambda}^{(l_1,p) (l_2,t)}
(\vec{k},\alpha; \vec{k},\alpha)$
are given by
\begin{eqnarray}
 & & c_{\Lambda}^{(l_1,p) (l_2,t)}(\vec{k},\alpha;\,\vec{k},\alpha) = 
  \sum_{\lambda_1(l_1)} \sum_{\lambda_2(l_2)} \sum_{s_z} 
   \nonumber \\
 & & \times 
  [\gamma A]_{\lambda_1}^{p\, *}(\vec{k},\alpha,s_z) \, 
   [\gamma A]_{\lambda_2}^t(\vec{k},\alpha,s_z) \,
    c_{\Lambda}(\lambda_1,\,\lambda_2) .
   \label{3.19}
\end{eqnarray}
We recall that $\lambda=(l,\Gamma,\nu,k)$ (Appendix A) and
the summations in (\ref{3.19}) are performed over the 
subindices $\Gamma,\nu,k$ of $\lambda(l_1)$ and $\lambda(l_2)$ 
within the manifolds $l_1$ and $l_2$.
Finally,
\begin{eqnarray}
 c_{\Lambda}(\lambda_1,\,\lambda_2)
 =\int S_{\lambda_1}(\Omega)\, 
 S_{\Lambda}(\Omega) \, S_{\lambda_2}(\Omega)\, d\Omega .  
\label{3.20}
\end{eqnarray}
These coefficients can be tabulated before
the self-consistent-field HFR procedure. 
The density of conduction electrons stays invariant under
all symmetry operations, which means that in expansion (\ref{3.20})
we consider only irreducible representations of $A_{1g}$ symmetry,
$\Gamma_{\Lambda}=A_{1g}$.
In such case the nonzero coefficients in (\ref{3.20}) can occur only if
(1) $\Gamma_{\lambda_1}=\Gamma_{\lambda_2}$ or
(2) both $\Gamma_{\lambda_1}$ and $\Gamma_{\lambda_2}$ belong
to the $A_{1g}$ irreducible representation. 

In the interstitial region $\rho_{val}(\vec{R})$ is expanded in Fourier
series,
\begin{eqnarray}
 \rho_{val}(\vec{R})=
 \sum_{\vec{K}} 
 \rho_{val}(\vec{K}) \, e^{i \vec{K} \vec{R}} ,
  \label{3.21}
\end{eqnarray}
where from Eqs.\ (\ref{b.1}) and (\ref{b.6}) we find
\begin{eqnarray}
 \rho_{val}(\vec{K}) & = & \frac{1}{N v} \sum_{\vec{k},\alpha}
 \sum_{\vec{K}'} \sum_{s_z} n_{\vec{k}, \alpha} \nonumber  \\
 & & \times  \gamma^*_{\vec{K}'}(\vec{k},\alpha,s_z) \,  
 \gamma_{\vec{K}+\vec{K}'}(\vec{k},\alpha,s_z) . 
\label{3.22} 
\end{eqnarray}

\subsection {Direct Coulomb matrix elements of core states} 

If one wants to use the Hartree-Fock-Roothaan method
then the problem of the radial dependence of core
states arises. In LDA it is solved quite naturally
since it is possible to introduce a self-consistent 
spherically symmetric potential
which includes an average exchange term. Nonspherical
contributions are small and usually omitted.
Then the radial components are obtained through the solution
of the Schr\"{o}dinger (or Dirac) 
equation in the potential.
However, there is no such convenience in the
Hartree-Fock approach since the exchange contribution 
generally can not be reduced to an effective single particle potential.
In the HF method one routinely uses radial dependencies of
Gaussian or Slater-type (GT or ST). 
An alternative choice of complete radial
basis functions is given in Appendix D. In all cases
the radial part is approximated as
\begin{eqnarray}
   {\cal R}_{\tau}(r)=\sum_{\eta} \gamma_{\tau,\eta} u^{\tau}_{\eta}(r) ,
 \label{4.1}
\end{eqnarray}
where $u_{\eta}$ is a radial basis function (GT, ST or other), 
while $\eta$ stands for
the parameters specifying the function.

The orientational 
vectors $\langle \hat{x} | j,l, m_j \rangle $ are usually
spherical harmonic spinors. \cite{Tin} 
In a crystal,  
the spherical symmetry is reduced, Eq.\ (\ref{3.24}), and we replace the
spinors by their symmetry adapted combinations 
$\langle \hat{x} | \tau \rangle $ as we discussed
in Sec.\ IV A1 and Appendix C. 
Then the basis function of a core state $\tau$ reads
\begin{eqnarray}
  \chi_{\eta}^{\tau}(x) \equiv \langle x | \tau, \eta \rangle 
  = u^{\tau}_{\eta}(r) \langle \hat{x} | \tau \rangle .
 \label{4.2}
\end{eqnarray}
Since the orientational vector $\langle \hat{x} | \tau \rangle$
is independent of $\eta$, the basis functions are distinguished only by
the radial component $u^{\tau}_{\eta}(r)$. 

In order to calculate the matrix elements of the direct Coulomb
interaction
for each atomic orbital $\tau$ we introduce the Hartree
potential so that no self-interaction occurs,
\begin{eqnarray}
   V^{\tau}(\vec{r})=V(\vec{R})-{\cal V'}^{\tau}(\vec{r}) =
   V^{\tau}_0(\vec{r})+V^{\tau}_M(\vec{r}) ,
 \label{4.3}
\end{eqnarray}
where $V^{\tau}_0$ and $V^{\tau}_M$ refer to
the spherical and the nonspherical components of $V^{\tau}$,
respectively.
First we consider the spherically symmetric component,
\begin{eqnarray}
   V^{\tau}_0(\vec{r})=V_0(\vec{R})-{\cal V'}^{\tau}_0(\vec{r}) .
 \label{4.4}
\end{eqnarray}
We get
\begin{eqnarray}
 \langle \tau, \eta | {\cal F}_0^{Coul} | \tau, \eta' \rangle &=& 
  \langle \tau, \eta | T | \tau, \eta' \rangle  
  \nonumber \\
  & + &   
 \int_0^R u_{\eta}^{\tau}(r) u_{\eta'}^{\tau}(r) V_0^{\tau}(r) r^2 dr .
 \label{4.5}
\end{eqnarray}
Here $\langle \tau, \eta | T | \tau', \eta' \rangle$ are the kinetic 
energy integrals which are well known for GT (ST) orbitals.
For the alternative set of basis radial functions (Appendix D)
instead of (\ref{4.5}) the matrix elements are 
given by Eq.\ (\ref{e.7}).

In order to compute the 
matrix elements of $V^{\tau}_M$
we partition $V(\vec{R})$ in two
parts, Eq.\ (\ref{2.20}). For $U_S$
(the potential of the charges inside the sphere $S(\vec{n})$) we
distinguish further two contributions, from the itinerant and
the core electrons, Eq.\ (\ref{3.14}).
Using equations derived in Sec.\ II, we obtain
\begin{eqnarray}
  \langle \tau, \eta | {\cal F}_M^{Coul} | \tau, \eta' \rangle &=& 
  {\sum_{\Lambda \neq 0}}' c_{\Lambda}(\tau,\tau)
   ( D_{\Lambda}^{\eta,\eta'}(out)
    \nonumber \\
   &+& D_{\Lambda}^{\eta,\eta'}(val)+
   D_{\Lambda}^{\eta,\eta'}(core) ) ,
 \label{4.6}
\end{eqnarray}
where 
\begin{eqnarray}
 D_{\Lambda}^{\eta,\eta'}(out)= 
 \frac{V_{\Lambda}^{out}}{R^l}
  q_{l}^{\eta \eta'}  .  
  \label{4.7}
\end{eqnarray}
Here $V_{\Lambda}^{out}$ is given by (\ref{2.23b}) and
the multipolar moments $q_{l}^{\eta \eta'}$ are 
\begin{eqnarray}
 q_{l}^{\eta \eta'}  =
 \int_0^R
 u_{\eta}^{\tau}(r) u_{\eta'}^{\tau}(r) r^{l+2} dr .
\label{4.8} 
\end{eqnarray}
For the contribution from the extended states, $D_{\Lambda}^{\eta,\eta'}(val)$, 
we obtain
\begin{eqnarray}
  D_{\Lambda}^{\eta,\eta'}(val) = 
 \sum_{l_1 l_2} C_{l}(u_{l_1}^p u_{l_2}^t| u_{\eta}^{\tau} u_{\eta'}^{\tau})\; 
  \rho_{\Lambda}^{(l_1,p) (l_2,t)} ,
 \label{4.9}
\end{eqnarray}
where the two-fold radial integrals $C_l(...)$ are defined by (\ref{d.1})
and $\rho_{\Lambda}^{(l_1,p) (l_2,t)}$ is defined by 
Eqs.\ (\ref{3.18}), (\ref{3.19}).
The contribution from the other core electrons reads
\begin{eqnarray}
  D_{\Lambda}^{\eta,\eta'}(core)= {\sum_{\tau'}}
  C_{l}(u_{\eta}^{\tau} u_{\eta'}^{\tau}| {\cal R}_{\tau'} {\cal R}_{\tau'}) \,
   c_{\Lambda}(\tau',\tau')\, n_{\tau'}^{\tau}  .
 \label{4.10}
\end{eqnarray}
The last sum runs over all occupied states 
{\it except} $\tau$ (no self-interaction), {\it i.e.} $n_{\tau}^{\tau}=0$
and $n_{\tau}^{\tau'}=1$ if the core state $\tau' \neq \tau$ 
is occupied and zero otherwise.

In cubic crystals only $f$ and $d$ shells are split by CEF effects,
Appendix~C. The CEF interaction of localized electrons 
with the delocalized ones
(second part of (\ref{4.6}))
has been discussed in a number of papers, 
Refs.\ \onlinecite{New,Ste,Bro,Nik2} and calculated
in Refs.~\onlinecite{Ste,Bro}.
The important result here is that
we have obtained {\it all CEF interactions}.
In particular, besides those considered in Refs.\ \onlinecite{Ste,Bro}, we
include the CEF effects from the rest of the crystal 
\cite{Nik1,Nik2,Nik3}
(first contribution, Eq.\ (\ref{4.6})) and from the other core shells
with nonspherical density (third part of (\ref{4.6})). 

\subsection{Direct Coulomb matrix elements of extended states}

Here we consider matrix elements of ${\cal F}^{Coul}$, Eq.\ (\ref{3.6}), 
for a conduction 
electron $|d \rangle \equiv | \vec{k}, \alpha \rangle$.
Usually in the LAPW method the basis functions 
$\chi_{\vec{k},\vec{K}}(\vec{R})=\langle \vec{R} | \vec{k}, \vec{K} \rangle$
(see Appendix B)
are defined in an effective potential 
$V_0^{LDA}=V_0+V_0^{exc}$ which includes the direct Coulomb
interaction $V_0$ and the
LDA exchange potential $V_0^{exc}$. \cite{And,Koe,Sin}
The potential $V_0^{LDA}$ is spherically symmetric
inside the ``muffin-tin" (MT) spheres and is a constant in the
interstitial region. Notice that $V_0^{LDA}$ is used only for
the construction of the basis functions
$\chi_{\vec{k},\vec{K}}(\vec{R})$, Eq.\ (\ref{b.1}). 
Having defined $\chi_{\vec{k},\vec{K}}(\vec{R})$, in principle
one can calculate the matrix elements of a general potential 
$V^{tot}(\vec{R})=V^{Coul}(\vec{R})+V^{exc}(\vec{R})$ 
(the procedure is known as FP-LAPW method \cite{Wim,Ham,Sin}). 
Below we follow the same approach,
but instead of $V_0^{LDA}$ for construction of the basis functions
we consider only the electrostatic potential $V_0^d(\vec{R})$ 
{\it without any exchange}, 
\begin{eqnarray}
  V_0^d(\vec{R})=V_0(\vec{R})-{\cal V'}_0^d(\vec{R}) .
 \label{4.11}
\end{eqnarray}
Here ${\cal V'}_0^d(\vec{R})$ is the spherically symmetric 
Coulomb potential due to the electron $d$. 
As we will see later in section V 
we can omit ${\cal V'}^d(\vec{R})$
for any conduction electron, 
because the corresponding matrix elements, Eq.\ (\ref{6.29}), 
decreases as
\begin{eqnarray}
   \langle \vec{k},\vec{K}' | {\cal V'}^d  | \vec{k},\vec{K} \rangle
   \sim \frac{1}{N} ,
 \label{4.12}
\end{eqnarray}
and vanish in the limit $N \rightarrow \infty$.
Therefore, constructing the LAPW basis functions
$\chi_{\vec{k},\vec{K}}$, we can use the potential $V_0(\vec{R})$,
{\it i.e.} $V_0^d(\vec{R})=V_0(\vec{R})$.
Next step is to calculate the direct matrix elements for the conduction
electron $d$ in the full Coulomb potential $V(\vec{R})$.
Following the method described in Sec.\ II we write 
$V(\vec{R})=V_0(\vec{R})+V_M(\vec{R})$
for the potential inside the spheres
($V_0$ and $V_M$ are the spherically symmetric part and the
contribution due to the other multipoles, correspondingly).
In the interstitial region we use the Fourier expansion of $V(\vec{R})$.
Then the Hartree-Fock operator ${\cal F}^{Coul}$ is separated into three parts,
\begin{eqnarray}
   {\cal F}^{Coul}={\cal F}_0^{Coul}+{\cal F}_M^{Coul}
   +{\cal F}_I^{Coul} ,
 \label{4.12'}
\end{eqnarray}
where ${\cal F}_0^{Coul}$ comprises the kinetic energy and the spherically
symmetric potential $V_0(\vec{R})$, $F_M^{Coul}$ accounts for
the other multipole terms while
${\cal F}_I^{Coul}$ stands for the electrostatic interaction in the
interstitial region.
Starting with ${\cal F}_0^{Coul}$ 
we arrive at the standard expressions for the matrix elements
of the LAPW method, \cite{Koe}
\begin{eqnarray}
 \langle \vec{p},\vec{K}',s'_z | {\cal F}_0^{Coul}+K_I | 
 \vec{k},\vec{K},s_z \rangle
 =\delta_{\vec{p},\vec{k}} \delta_{s_z s'_z}
  H_{\vec{K}',\vec{K}}(\vec{k})
 \label{4.14}
\end{eqnarray}
where $H_{\vec{K}',\vec{K}}(\vec{k})$ is given by
Eq.\ (15) of Ref.~\onlinecite{Koe}. (The overlap matrix is
given by Eq.\ (13a) of Ref.~\onlinecite{Koe}.)

The other contributions of the general Coulomb potential $V^{Coul}(\vec{R})$ 
follow from equations of Sec.\ II.
After some algebra, we obtain:
\begin{eqnarray}
 & & \langle \vec{p},\vec{K}',s'_z |{\cal F}_M^{Coul}+{\cal F}_I^{Coul}
 | \vec{k},\vec{K},s_z \rangle  
      =  \delta_{\vec{k} \vec{p}}  \delta_{s_z s'_z}   \label{4.15} \\ 
 & & \times  \left(
   {\sum_{\Lambda \neq 0} }' \sum_{l_1 l_2} 
     B_{\Lambda}^{(l_1,p)(l_2,t)} 
   \bar{c}_{\Lambda}^{(l_1,p)(l_2,t)}  
     + B_I \right) . 
   \nonumber
\end{eqnarray}
Here $B_I$ arises due to the interstitial contribution,
\begin{eqnarray}
 B_I={\sum_{\vec{P}\neq 0}}' V(\vec{P})\,{\cal O}(\vec{K}-\vec{K}'+\vec{P}) ,
 \label{4.15a}
\end{eqnarray}
where $V(\vec{P})$ stands for the Fourier coefficients, Eq.\ (\ref{2.17b}).
${\cal O}(\vec{P}')$ is given by (\ref{7.4b}) and
\begin{eqnarray}
 & & \bar{c}_{\Lambda}^{(l_1,p)(l_2,t)}=\sum_{\lambda_1(l_1)} 
 \sum_{\lambda_2(l_2)}
   A_{\lambda_1}^{p*}(\vec{k},\vec{K}') A_{\lambda_2}^t(\vec{k},\vec{K})\,
   c_{\Lambda}(\lambda_1,\lambda_2) . \nonumber \\
 & & \label{4.16}
\end{eqnarray}
To condense notations we shall use $L$ for $(l,p)$. In (\ref{4.15})
\begin{eqnarray}
   B_{\Lambda}^{L_1 L_2}
 =  B_{\Lambda}^{L_1 L_2}(out) +
   B_{\Lambda}^{L_1 L_2}(val)
 +  B_{\Lambda}^{L_1 L_2}(core) ,
 \label{4.17}
\end{eqnarray}
where the index $out$ stands for the contribution from 
the other spheres,
\begin{eqnarray}
   B_{\Lambda}^{L_1 L_2}(out)= 
 \frac{V_{\Lambda}^{out}}{R^l}
  q_{l}^{L_1 L_2} .
 \label{4.18}
\end{eqnarray}
Here $V_{\Lambda}^{out}$ is given by (\ref{2.23b})
and the multipolar moments $q_{\Lambda}^{L_1 L_2}$ are  
\begin{eqnarray}
 q_{l}^{(l_1,p)(l_2,t)}=
 \int_0^R
 u_{l_1}^p(r) u_{l_2}^t(r) r^{l+2} dr .
\label{4.19} 
\end{eqnarray}
For the contribution from the delocalized electrons we obtain
\begin{eqnarray}
   B_{\Lambda}^{(l_1,p)(l_2,t)}(val)= 
 \sum_{l'_1 l'_2} 
 C_l(u_{l_1}^p u_{l_2}^t | u_{l'_1}^{p'} u_{l'_2}^{t'}) \; 
  \rho_{\Lambda}^{(l'_1,p')(l'_2,t')} ,
 \label{4.20}
\end{eqnarray}
where $\rho_{\Lambda}^{L_1 L_2}$ and 
$C_l(...)$ are given by Eqs.\ (\ref{3.18}) and (\ref{d.1}),
respectively.

For the contribution from the closed shell core electrons we have
\begin{eqnarray}
   B_{\Lambda}^{(l_1,p)(l_2,t)}(core)= \sum_{\tau} 
  C_{l}(u_{l_1}^p u_{l_2}^t | {\cal R}_{\tau} {\cal R}_{\tau}) \,
   c_{\Lambda}(\tau,\tau)\, n_{\tau}  ,
 \label{4.21}
\end{eqnarray}
where again the two-fold radial integral $C_l$  
is defined by Eq.\ (\ref{d.1}).

The non-spherical components of the total Coulomb potential in (\ref{4.15})
represent crystal field like effects for conduction electrons.

\section {Calculations of the Exchange Matrix Elements} 
\label{sec:Exc}

In this section we calculate the exchange matrix elements 
of the Hartree-Fock method. 
The most general expression for exchange is \cite{Roo}
\begin{eqnarray}
& & \langle a | {\cal F}^{exc} | b \rangle =
 \sum_c \int \chi_a^*(x_1) \psi_c(x_1) \frac{1}{R_{12}} 
 \chi_b(x_2) \psi_c^*(x_2) dx_1 dx_2 . \nonumber \\
 & & \label{5.1}
\end{eqnarray}
Here both indices $a$ and $b$ refer to basis wave
functions of an electronic state $d$ (which can be either 
a conduction or a localized state), Eq.\ (\ref{3.3});
$\psi_c$ stand for the estimated wave functions
of conduction and localized electrons obtained from a
previous iteration of the HFR self-consistent procedure.
The summation is understood over all occupied states $c$.
As before (Sec.\ III), $x=(\vec{R},s_z)$ and
the integration (\ref{5.1}) includes summation over two spin
components $s_z$.
We will calculate the matrix element (\ref{5.1}) in two steps:
1) we construct an auxiliary Coulomb potential
\begin{eqnarray}
 V_{cb}(\vec{R}) = \int 
 \frac{\rho_{cb}(\vec{R}')}{|\vec{R}-\vec{R}'|} 
   d\vec{R}' ,
  \label{5.3}
\end{eqnarray}
which corresponds to the ``exchange" density
\begin{eqnarray}
 \rho_{cb}(\vec{R}) =  \sum_{s_z=-1/2}^{1/2}
 \psi_c^*(\vec{R},s_z) \chi_b(\vec{R},s_z)  .
  \label{5.4}
\end{eqnarray}
($\psi_c$ has two spin components due to the spin-orbit
interaction, Eq.\ (\ref{3.5a}).);
2) We calculate the matrix element of exchange as
\begin{eqnarray}
  \langle a | {\cal F}^{exc} | b \rangle = \sum_c
  \int V_{cb}(\vec{R}) \rho_{ca}^*(\vec{R}) d\vec{R} .
  \label{5.5} 
\end{eqnarray}
We want to stress that the ``exchange"
potential $V_{cb}$ and the ``exchange" density $\rho_{cb}$
are technical quantities here, which are employed only for the calculation of
the matrix element of exchange, Eq.\ (\ref{5.5}).
They should not be confused with the effective exchange potential which
is derived and widely used in the local density approximation.

\subsection{Exchange for localized electrons}

Here we calculate the exchange for a localized state $\tau$ sited at $\vec{n}$,
$i.e.$ $d \equiv \tau$.
Index $c$ can refer either
to a conduction state or to another core state $\tau'$.
First we consider $c$ as a conduction state, $i.e.$ 
$| c \rangle \equiv | \vec{k},\alpha \rangle$. 
The radial wave function of the localized electron $\tau$
is expanded in terms of $u_{\eta}^{\tau}$, 
Eq.\ (\ref{4.1}), so that $| a \rangle$ and $| b \rangle$
stand for $| \tau, \eta \rangle$. 
Since the state $\tau$ is confined inside $S(\vec{n})$,
for the calculation of the exchange (\ref{5.5}) 
we need to know the density $\rho_{cb}$ and
the potential $V_{cb}$, Eqs.\ (\ref{5.4}) and (\ref{5.3}), 
only inside the same sphere.
For $\rho_{cb}$ we get
\begin{eqnarray}
& &  \rho_{(\vec{k},\alpha)\, (\tau,\eta)}(\vec{r}(\vec{n}))
  = 
 \frac{1}{\sqrt{N}}  \sum_{\Lambda} 
 \rho^{(\vec{k},\alpha)\, \tau}_{\Lambda}
 (r(\vec{n})) \, S_{\Lambda}(\hat{r}(\vec{n})) , \nonumber \\
& & \label{5.6}
\end{eqnarray}
where 
\begin{eqnarray}
 \rho^{(\vec{k},\alpha)\, (\tau,\eta)}_{\Lambda}(r) 
 & = &  \sum_{\lambda'} \sum_{s_z}
 c_{\Lambda}(\tau,s_z;\lambda')\, 
 [\gamma A]_{\lambda'}^{p\,*}(\vec{k},\alpha,s_z)\, \nonumber \\
 & & \times u_{l'}^p(r) u_{\eta}^{\tau}(r) .
 \label{5.7}
\end{eqnarray}
The coefficients $c_{\Lambda}(\tau,s_z;\lambda')$ are given by
\begin{eqnarray}
 c_{\Lambda}(\tau,s_z;\lambda)=
 \int \langle \Omega,s_z | \tau \rangle  S_{\Lambda}(\Omega) 
  S_{\lambda}(\Omega) d\Omega .
 \label{5.7b}
\end{eqnarray}
From (\ref{5.6}) the exchange between $|\tau \rangle$ and 
$| \vec{k},\alpha \rangle$ is found as
\begin{eqnarray}
 \langle \tau, \eta | {\cal F}^{exc}(\vec{k},\alpha) | \tau, \eta' \rangle
 =\frac{1}{N} D^{exc}(\vec{k},\alpha) ,
  \label{5.8} 
\end{eqnarray}
where
\begin{eqnarray}
  & & D^{exc}(\vec{k},\alpha)  = 
  \sum_{\Lambda}  \sum_{\lambda_1 \lambda_2} \sum_{s_z,s'_z}
  c_{\Lambda}(\tau,s_z;\lambda_1)\, c_{\Lambda}(\tau,s'_z;\lambda_2) 
  \nonumber \\
  & & \times [\gamma A]_{\lambda_1}^{p\,*}(\vec{k},\alpha,s_z)\,
  [\gamma A]_{\lambda_2}^{p'}(\vec{k},\alpha,s'_z)\,
  C_l(u_{l_1}^p u_{\eta}^{\tau} | u_{l_2}^{p'} u_{\eta'}^{\tau}) .
  \label{5.9} 
\end{eqnarray}
Here the integrals $C_l(...)$ are given by Eq.\ (\ref{d.1}).
The exchange with all extended states reads
\begin{eqnarray}
 \langle \tau, \eta | {\cal F}^{exc}(c) | \tau, \eta' \rangle
 =\frac{1}{N} \sum_{\vec{k},\alpha} D^{exc}(\vec{k},\alpha) \,
 n_{\vec{k},\alpha}.
  \label{5.10} 
\end{eqnarray}

Next we consider the exchange (\ref{5.5}) between $\tau$
(as before, $\tau \equiv d$) and a core electron
$\tau'$ localized at the same site $\vec{n}$.
$\tau'$ is now our reference state $c$, {\it i.e.} $c=\tau'$.
Proceeding analogously, we find that
the multipole expansion of the ``exchange" density $\rho_{cb}$ is 
\begin{eqnarray}
 \rho_{\tau'\, (\tau,\eta)}(\vec{r}(\vec{n}))
  = 
   \sum_{\Lambda} 
 \rho^{\tau'\, (\tau,\eta)}_{\Lambda}
 (r(\vec{n})) \, S_{\Lambda}(\hat{r}(\vec{n})) ,   
 \label{5.11}
\end{eqnarray}
where 
\begin{eqnarray}
 \rho^{\tau'\, (\tau,\eta)}_{\Lambda}(r) 
  =   
 c_{\Lambda}(\tau,\tau') {\cal R}_{\tau'}(r) u_{\eta}^{\tau}(r) .
 \label{5.12}
\end{eqnarray}
The exchange integral is
\begin{mathletters}
\begin{eqnarray}
 \langle \tau, \eta | {\cal F}^{exc}(core) | \tau, \eta' \rangle 
 = \sum_{\tau'} n_{\tau}^{\tau'} D^{exc}(\tau') ,
         \label{5.12'a}  
\end{eqnarray}
where
\begin{eqnarray}
  D^{exc}(\tau')=\sum_{\Lambda} 
 | c_{\Lambda}(\tau,\tau') |^2\,
 C_l( {\cal R}_{\tau'} u_{\eta}^{\tau} |
  {\cal R}_{\tau'} u_{\eta'}^{\tau} ) .
 \label{5.12'b}
\end{eqnarray}
\end{mathletters}

\subsection{Exchange between an extended state with the localized states}

In this subsection
we consider $d$ as an extended electron state, 
{\it i.e.} $d=(\vec{k},\alpha)$,
while $c$ refers to a core state $\tau$ 
localized inside a sphere $S(\vec{n})$, $c=\tau(\vec{n})$.
We expand the wave function $\langle \vec{R} | \vec{k},\alpha \rangle$
in terms of $\langle \vec{R}, s_z | \vec{k}, \vec{K} \rangle =
\langle \vec{R} |  \vec{k}, \vec{K} \rangle \zeta(s_z)$,
Appendix B. Therefore, $a,b=(\vec{k},\vec{K},s_z)$.
The exchange density $\rho_{c b}$, Eq.\ (\ref{5.4}),
is located inside the sphere $S(\vec{n})$,
\begin{eqnarray}
  \rho_{\tau\, (\vec{k},\vec{K},s_z)}(\vec{r})=\frac{1}{\sqrt{N}} 
  e^{i \vec{k}\cdot \vec{X}(\vec{n})}
  \sum_{\Lambda} \sum_{\lambda'}  
  \left( {\cal R}_{\tau}(r) u_{l'}^p(r) \right.
  \nonumber \\
  \left. 
  \times c_{\Lambda}(\tau,s_z;\lambda') A_{\lambda'}^p(\vec{k},\vec{K})
  \right) S_{\Lambda}(\hat{r}) ,
  \label{6.1} 
\end{eqnarray}
where the coefficients $c_{\Lambda}(\tau,s_z;\lambda')$ are given
by Eq.\ (\ref{5.7b}). Again, using the multipole expansion (\ref{6.1})
we calculate the exchange (\ref{5.5}) and obtain
\begin{eqnarray}
 \langle \vec{k},\vec{K},s_z | {\cal F}^{exc}(\tau(\vec{n})) |
 \vec{k},\vec{K}',s'_z \rangle   
  = \frac{1}{N} B^{exc}(\tau) , 
  \label{6.2}
\end{eqnarray}
where
\begin{eqnarray}
  & & B^{exc}(\tau) = 
  \sum_{\Lambda} \sum_{\lambda_1 \lambda_2} 
 c_{\Lambda}(\tau,s_z;\lambda_1)\, c_{\Lambda}(\tau,s'_z;\lambda_2)
   \nonumber \\
 & &  \times   
 A_{\lambda_1}^{p\,*}(\vec{k},\vec{K}) A_{\lambda_2}^t(\vec{k},\vec{K}')\,
 C_{l}({\cal R}_{\tau} u_{l_1}^p | {\cal R}_{\tau} u_{l_2}^t ) .
  \label{6.3}
\end{eqnarray}
Notice that the latter result is independent of $\vec{n}$ and
the matrix elements (\ref{6.2}) can be nonzero even for off-diagonal
spin functions ($s_z=-1/2$, $s'_z=1/2$ or vice versa) due to
the spin-orbit interaction.
In order to obtain the exchange with all localized electrons, we sum
(\ref{6.2}) over all sites $\vec{n}$ and all occupied core
electron states $| \tau \rangle$, and find that
\begin{eqnarray}
 \langle \vec{k},\vec{K},s_z | {\cal F}^{exc}(core) | 
 \vec{k},\vec{K}',s'_z \rangle  
 = \sum_{\tau} 
   B^{exc}(\tau)  .
  \label{6.4}
\end{eqnarray}

\subsection{Exchange between extended states}

The calculation of exchange between two delocalized electrons is much more 
involved and quite laborious. 
As the state $c$ in Eqs.\ (\ref{5.1})-(\ref{5.5})
we consider now a conduction state $| \vec{p},\beta \rangle$
with the wave function 
$\psi_c(\vec{R})=\langle x | \vec{p},\beta \rangle$.
As before, $d=(\vec{k},\alpha)$ and the wave function of the $d$th
electron, $\langle \vec{R} | \vec{k},\alpha \rangle$, is expanded in
terms of LAPW basis functions $\langle \vec{R}, s_z | \vec{k}, \vec{K} \rangle =
\langle \vec{R} |  \vec{k}, \vec{K} \rangle \zeta(s_z)$ (Appendix~B)
labeled by indices $a,b=(\vec{k},\vec{K},s_z)$.
In this subsection we first calculate the matrix elements of $d$th electron
exchanged with the extended state $c$. Then by summing over all
occupied extended states $|c \rangle \equiv | \vec{p},\beta \rangle$ 
we will be able to compute the matrix element of exchange between $d$ 
and the other conduction electrons.

Inside a sphere $S(\vec{n})$ the ``exchange" density 
$\rho_{cb}$, Eq.\ (\ref{5.4}), is given by
\begin{eqnarray}
& &  \rho_{c\,(\vec{k},\vec{K},s_z) }(\vec{r})
  = 
 \frac{1}{N} e^{i \vec{q} \vec{X}(\vec{n})} \sum_{\Lambda} 
 \rho^{c\,(\vec{k},\vec{K},s_z)}_{\Lambda}
 (r(\vec{n})) \, S_{\Lambda}(\hat{r}(\vec{n})) , \nonumber \\
& & \label{6.5}
\end{eqnarray}
where $\vec{q}=\vec{p}-\vec{k}$ and
\begin{eqnarray}
 \rho^{c\,(\vec{k},\vec{K},s_z)}_{\Lambda}(r) & = & \sum_{l_1 l_2}
 \sum_{\lambda_1 \lambda_2}
 c_{\Lambda}(\lambda_1,\lambda_2)\, 
 [\gamma A]_{\lambda_1}^{p\,*}(\vec{p},\beta,s_z)
 \nonumber \\
 & & \times A_{\lambda_2}^t(\vec{k},\vec{K}) \,
 u_{l_1}^p(r) u_{l_2}^t(r) .
 \label{6.6}
\end{eqnarray}
Outside the spheres $\rho_{c b}$
is expanded in plane waves:
\begin{eqnarray}
 \rho_{c\,(\vec{k},\vec{K},s_z)}(\vec{R})=\frac{1}{N v}
 e^{i \vec{q} \cdot \vec{R}}   \sum_{\vec{K}'} 
  \gamma^*_{\vec{K}'+\vec{K}}(\vec{p},\beta,s_z) \, 
  e^{i \vec{K}' \cdot \vec{R}} .
  \label{6.7}
\end{eqnarray}

Proceeding
as in Sec.\ II we continue the plane wave representation (\ref{6.7})
inside the spheres and then subtract it out from Eq.\ (\ref{6.5}).
This procedure renormalizes the multipole radial functions  
$\rho^{c\,(\vec{k},\vec{K},s_z)}_{\Lambda}$ of (\ref{6.6}),
which now are given by
\begin{eqnarray}
 & & 
 {\rho'}^{\,c\,(\vec{k},\vec{K},s_z)}_{\Lambda}(r) 
 = \rho^{c\,(\vec{k},\vec{K},s_z)}_{\Lambda}(r)  - \frac{4\pi i^l}{v}
 \nonumber \\
 & &  \times
 \sum_{\vec{K}'} j_l(|\vec{K}'+\vec{q}|r)\, S_{\Lambda}(\vec{K}'+\vec{q})\,
  \gamma^*_{\vec{K}'+\vec{K}}(\vec{p},\beta,s_z) .
  \label{6.8}
\end{eqnarray}
Here $S_{\Lambda}(\vec{K}'+\vec{q})$ is understood as the symmetry adapted
function $S_{\Lambda}(\Omega_{\vec{K}'+\vec{q}})$ of the polar angles 
$\Omega_{\vec{K}'+\vec{q}}$ defined by the vector $\vec{K}'+\vec{q}$.

Notice that $\rho_{c\,(\vec{k},\vec{K},s_z)}$ is not a periodic 
function of $\vec{R}$. A translation by $\vec{t}=\vec{X}(\vec{n})$
transforms it as
\begin{mathletters}
\begin{eqnarray}
  \vec{t}\;\rho_{c\,(\vec{k},\vec{K},s_z)}(\vec{R})=
  e^{i \vec{q} \vec{X}(\vec{n})} \rho_{c\,(\vec{k},\vec{K},s_z)}(\vec{R}) .
  \label{6.7b}
\end{eqnarray}
This transformation is not identical for $\vec{q} \neq 0$.
From Eq.\ (\ref{5.3}) we find that the same transformational law 
holds for the corresponding potential, {\it i.e.}
\begin{eqnarray}
  \vec{t}\; V_{c\,(\vec{k},\vec{K},s_z)}(\vec{R})=
  e^{i \vec{q} \vec{X}(\vec{n})} V_{c\,(\vec{k},\vec{K},s_z)}(\vec{R}) .
  \label{6.7c}
\end{eqnarray}
\end{mathletters}
As a consequence of Bloch's theorem we obtain
\begin{mathletters}
\begin{eqnarray}
   & & \rho_{c\,(\vec{k},\vec{K},s_z)}(\vec{R})=
   e^{i \vec{q} \vec{R}} \tilde{\rho}_{c\,(\vec{k},\vec{K},s_z)}(\vec{R}) ,
  \label{6.7d} \\
   & & V_{c\,(\vec{k},\vec{K},s_z)}(\vec{R})=
   e^{i \vec{q} \vec{R}} \tilde{V}_{c\,(\vec{k},\vec{K},s_z)}(\vec{R}) ,
  \label{6.7e}  
\end{eqnarray}
\end{mathletters}
where $\tilde{\rho}_{c\,b}$ and $\tilde{V}_{c\,b}$ are periodic functions, 
{\it i.e.}
\begin{mathletters}
\begin{eqnarray}
   & & \vec{t}\;\tilde{\rho}_{c\,(\vec{k},\vec{K},s_z)}(\vec{R})=
   \tilde{\rho}_{c\,(\vec{k},\vec{K},s_z)}(\vec{R}) ,
  \label{6.7f} \\
   & & \vec{t}\;\tilde{V}_{c\,(\vec{k},\vec{K},s_z)}(\vec{R})=
    \tilde{V}_{c\,(\vec{k},\vec{K},s_z)}(\vec{R}) .
  \label{6.7g}  
\end{eqnarray}
\end{mathletters}
Therefore, in the interstitial region
 $\tilde{\rho}_{c\,b}$ and $\tilde{V}_{c\,b}$
are expanded in Fourier series, while the initial functions
${\rho}_{c\,b}$ and ${V}_{c\,b}$ are expanded in terms of
$\exp(i (\vec{q}+\vec{K}') \vec{R})$, where $\vec{K}'$ is
a reciprocal lattice vector. 
The factor $\exp(i \vec{q} \vec{R})$ 
or, more precisely, $\exp(i \vec{q} \vec{X})$, will be also present
for solutions inside the spheres.
Indeed, if one knows a solution inside a sphere, say $S(\vec{n}=0)$
at the origin, then by means of Eq.\ (\ref{6.7c}) it is easy to
generate the solution inside any other sphere.
Notice, however, that the
factor $\exp(i \vec{q} \vec{X}(\vec{n}))$ cancels in the final
expression (\ref{5.5}).
As a result, it is easily to generalize the method of Sec.\ II for
the present consideration.

First we consider the potential in the interstitial region,
\begin{eqnarray}
 V_e(\vec{R})=\frac{1}{N} e^{i \vec{q}\vec{R}} 
 \sum_{\vec{K}'} 
 V_e(\vec{K}') \, 
  e^{i \vec{K}' \vec{R}} .
  \label{6.9}
\end{eqnarray}
As in Sec.\ II we distinguish three contributions there,
\begin{eqnarray}
 V_e(\vec{K}')  = V_e^I(\vec{K}')+V_e^{S,0}(\vec{K}')+V_e^{S,M}(\vec{K}') ,
  \label{6.10}
\end{eqnarray}
where $V_e^{S,0}(\vec{K}')$ and $V_e^{S,M}(\vec{K}')$
are the Fourier components of the potentials of the monopole and 
the other higher multipoles of the sphere, respectively.
$V_e^I(\vec{K}')$ represents a component from the 
the plane wave expansion (\ref{6.7}), {\it i.e.}
\begin{eqnarray}
 V_e^I(\vec{K}')= \frac{1}{v}
 \frac{4\pi}{|\vec{K}'+\vec{q}|^2} 
 \gamma^*_{\vec{K}'+\vec{K}}(\vec{p},\beta,s_z) .
  \label{6.11}
\end{eqnarray}
The Fourier components $V_e^{S,0}(\vec{K}')$ and $V_e^{S,M}(\vec{K}')$ 
are found as
\begin{eqnarray}
  V_e^{S,g}(\vec{K}') =\frac{1}{v} \int_{cell}
 e^{-i (\vec{q}+\vec{K}') \vec{R}}\, V_{S_g}(\vec{R})\, d^3R ,
  \label{6.9b}
\end{eqnarray}
where $g=0$, $M$, and
\begin{eqnarray}
 V_{S_g}(\vec{R}) = \sum_{\vec{n}} e^{i \vec{q} \vec{X}(\vec{n})}\,
 {\cal U}_{S_g}(\vec{R}-\vec{X}(\vec{n})) .
  \label{6.9c}
\end{eqnarray}
Here ${\cal U}_{S_0}$ and ${\cal U}_{S_M}$ are the 
potentials due to the monopole ($l=0$) and
the other multipoles ($l \ge 1$) of a single sphere (see Eq.\ (\ref{2.8})).
For the potential ${\cal U}_S$, Eq.\ (\ref{2.8}), we use
the equivalent charge distribution given by Eq.\ (\ref{6.8}).
The integral (\ref{6.9b}) is taken over a unit cell of the Bravais
lattice. One can show that $V_e^{S,g}(\vec{K}')$ is also expressed as \cite{Ewa1}
\begin{eqnarray}
  V_e^{S,g}(\vec{K}') =\frac{1}{v} \int_{crystal}
 e^{-i (\vec{q}+\vec{K}') \vec{R}}\, {\cal U}_{S_g}(\vec{R})\, d^3R ,
  \label{6.9d}
\end{eqnarray}
where the integration spans the whole crystal.
The representation (\ref{6.9d}) is then used to compute the Ewald
expansion coefficients $V_e^{S,0}(\vec{K}')$ and to find
$V_e^{S,M}(\vec{K}')$ by direct integration.
The procedure is the same as in Sec.\ II.

In order to proceed with the Ewald expansion,
we introduce an effective point ``exchange charge" inside the spheres,
\begin{eqnarray}
 Q^{c\,(\vec{k},\vec{K},s_z)}_0 = \sqrt{4 \pi} 
 \int_0^R {\rho'}^{\,c\,(\vec{k},\vec{K},s_z)}_0(r) r^2 . 
 \label{6.12}
\end{eqnarray}
Using the orthogonality of the radial functions $u_l^1$ and $u_l^2$ 
and the relation 
$c_0(\lambda,\lambda')=\delta_{\lambda \lambda'}/\sqrt{4 \pi}$,
we obtain
\begin{eqnarray}
  & & Q^{c\,(\vec{k},\vec{K},s_z)}_0 = \sum_{l} \sum_{\lambda} 
 [\gamma A]_{\lambda}^{p\, *}(\vec{p},\beta,s_z) \,
  A_{\lambda}^p(\vec{k},\vec{K}) \,
 N_l^p  \nonumber \\
 & & - \frac{4\pi R^2}{v}
 \sum_{\vec{K}'} \frac{j_1(|\vec{K}'+\vec{q}|R)}{|\vec{K}'+\vec{q}|} 
  \gamma^*_{\vec{K}'+\vec{K}}(\vec{p},\beta,s_z) ,
 \label{6.12'} 
\end{eqnarray}
where $N_l^1=1$ (normalization of the radial functions of the LAPW method) 
and
\begin{eqnarray}
  N_l^2 & = & \int_0^R [u_l^2(r)]^2 r^2 dr . \label{6.14}
\end{eqnarray}
Although in general $Q^{c\,(\vec{k},\vec{K},s_z)}_0(R) \neq 0$
in Appendix F we prove that for the present 
particular case (one atom in the unit cell)
this ``charge" has to be considered
only for $\vec{q}=0$, {\it i.e.} 
when $\vec{p}=\vec{q}$ but $\alpha \neq \beta$.
However, the final results will not be affected if one assumes that
$Q^{c\,(\vec{k},\vec{K},s_z)}_0(R) \neq 0$ for $\vec{q} \neq 0$.
If there are two or more atoms in the unit cell,
the ``exchange charges" of nonequivalent atoms must be 
introduced and taken into account
even for the case $\vec{p} \neq \vec{q}$, Appendix F.
We shall now proceed further as for the general case.
The Ewald expansion coefficients are given by
\begin{eqnarray}
   V_e^{S,0}(\vec{K}') =  \frac{\pi}{v} \frac{4}{|\vec{K}'+\vec{q}|^2}
   e^{-|\vec{K}'+\vec{q}|^2/4G^2} Q^{c\,(\vec{k},\vec{K},s_z)}_0 .
 \label{6.15}
\end{eqnarray}

For the higher multipoles we introduce two functions,
\begin{mathletters}
\begin{eqnarray}
 & & Q_{\Lambda}^{c\,(\vec{k},\vec{K},s_z)}(r)= 
  \int_0^r 
  {\rho'}_{\Lambda}^{\,c\,(\vec{k},\vec{K},s_z)}(r')\, {r'}^{\, l+2} dr' ,
  \label{f.16a} \\
 & & {Q'}_{\Lambda}^{\, c\,(\vec{k},\vec{K},s_z)}(r)= 
       \int_r^R {\rho'}_{\Lambda}^{\, c\,(\vec{k},\vec{K},s_z)}(r')\, 
       {r'}^{\,1-l} dr' .
 \label{6.16b} 
\end{eqnarray}
\end{mathletters}
Proceeding then as in section II we find:
\begin{eqnarray}
 & & V_e^{S,M}(\vec{K}') = \frac{(4\pi)^2}{v}  {\sum_{\Lambda \neq 0} }'
 \frac{(-i)^l}{2l+1} S_{\Lambda}(\vec{K}'+\vec{q}) \nonumber \\
 & & \times
 \left(  A_{\Lambda}(|\vec{K}'+\vec{q}|) + 
 Q_{\Lambda}^{c\,(\vec{k},\vec{K},s_z)}(R) \frac{j_{l-1}(|\vec{K}'+\vec{q}|R)}
 {|\vec{K}'+\vec{q}| R^{l-1}} \right) ,
 \label{6.17}  
\end{eqnarray}
where
\begin{eqnarray}
 A_{\Lambda}(|\vec{K}'+\vec{q}|) & = &
 \int_0^R \left( \frac{Q_{\Lambda}^{c\,(\vec{k},\vec{K},s_z)}(r)}{r^{l+1}} + 
 r^l {Q'}^{\,c\,(\vec{k},\vec{K},s_z)}_{\Lambda}(r) \right)  \nonumber \\
 & & \times j_l(|\vec{K}'+\vec{q}|r) r^2 dr .
 \label{6.18}  
\end{eqnarray}
Eqs.\ (\ref{6.11}), (\ref{6.15}) and (\ref{6.17}) fully determine
the plane wave components $V_e(\vec{K}')$, Eq.\ (\ref{6.10}).
Thus, we have obtained the potential $V_e(\vec{R})$, Eq.\ (\ref{6.9}), in the
interstitial region.

The potential inside a sphere $S(\vec{n})$ is found as
\begin{eqnarray}
  V_e(\vec{r})
 =\frac{1}{N} e^{i \vec{q} \vec{X}(\vec{n})} \, [ V_{e,0}(r) +
 {\sum_{\Lambda \neq 0}}' V_{e,\, \Lambda}(r)\, S_{\Lambda}(\hat{r} ) ] ,
 \label{6.19}  
\end{eqnarray}
where $V_{e,0}(r)$ and $V_{e,\, \Lambda}(r)$
are smooth functions of $r$. As before, we introduce
the constants $V_{e,0} \equiv V_{e,0}(R)$ and
$V_{e,\, \Lambda} \equiv V_{e,\, \Lambda}(R)$. 
They are found from the boundary-value problem for the surface of the sphere:
\begin{mathletters}
\begin{eqnarray}
 & & V_{e,\, 0} = \sum_{\vec{K}'}
 j_0(|\vec{K}'+\vec{q}|R)\, V_e(\vec{K}') ,  
 \label{6.20a} \\
 & & V_{e,\, \Lambda} = 4\pi i^l \sum_{\vec{K}'}
 j_l(|\vec{K}'+\vec{q}|R)\, S_{\Lambda}(\vec{K}'+\vec{q})\,
 V_e(\vec{K}') . \nonumber \\
 & & \label{6.20}  
\end{eqnarray}
\end{mathletters}
Here $V_{e,\, 0}$ and $V_{e,\, \Lambda}$ depend on $\vec{q}$
and have to be recalculated for each extended state $\vec{p}=\vec{k}+\vec{q}$.
We recall that $V_e(\vec{K}')$ is given by the sum (\ref{6.10}).
Following Sec.\ II.C we rewrite the potential $V_e(\vec{r})$, Eq.\ (\ref{6.19}), 
as
\begin{eqnarray}
  V_e(\vec{r}(\vec{n})) = \frac{1}{N} e^{i \vec{q} \vec{X}(\vec{n})} \,
 \left[ U_{e,S}(\vec{r}(\vec{n})) + V_{e,S}^{out}(\vec{r}(\vec{n})) \right] , 
 \label{6.20b}
\end{eqnarray}
where $U_{e,S}(\vec{r}(\vec{n}))$ is the potential of the
charges located inside the sphere $S(\vec{n})$ and
$V_{e,S}^{out}(\vec{r})$ accounts for the 
potential of the rest of the crystal. The single sphere
potential $U_{e,S}(\vec{r})$ is given by Eq.\ (\ref{2.21})
where $q_0(r)$, $q'_0(r)$ are replaced by
\begin{mathletters}
\begin{eqnarray}
 & & q_0^{\, c\,(\vec{k},\vec{K},s_z)} (r)= 
 \sqrt{4 \pi} \int_0^r 
 \rho_0^{\, c\,(\vec{k},\vec{K},s_z)} (r')\, {r'}^2 dr' ,
  \label{6.21a} \\
 & & {q'}_0^{\, c\,(\vec{k},\vec{K},s_z)} (r)=\sqrt{4 \pi} \int_r^R 
 \rho_0^{\, c\,(\vec{k},\vec{K},s_z)} (r')\, r' dr' , 
 \label{6.21b} 
\end{eqnarray}
and $q_{\Lambda}(r)$, $q_{\Lambda'}(r)$ by
\begin{eqnarray}
 & & q_{\Lambda}^{c\,(\vec{k},\vec{K},s_z)}(r)= 
  \int_0^r \rho_{\Lambda}^{\, c\,(\vec{k},\vec{K},s_z)}(r')\, 
  {r'}^{\, l+2} dr' ,
  \label{6.21c}  \\
 & & {q'}_{\Lambda}^{c\,(\vec{k},\vec{K},s_z)}(r)= 
  \int_r^R \rho_{\Lambda}^{\, c\,(\vec{k},\vec{K},s_z)}(r')\, 
  {r'}^{\, 1-l} dr' .  
  \label{6.21d}
\end{eqnarray}
\end{mathletters}
The potential of the ``exchange charges" outside the sphere
$S(\vec{n})$ is found by subtracting from Eq.\ (\ref{6.20a},b) 
the potential of the ``charges" inside. Thus, we introduce
the constants
\begin{mathletters}
\begin{eqnarray}
 & & V_{e,\, 0}^{out}=V_{e,\, 0}
  -\frac{q_0^{c\,(\vec{k},\vec{K},s_z)}(R)}{R} ,
  \label{6.21e} \\
 & & V_{e,\, \Lambda}^{out} =V_{e,\, \Lambda} -\frac{4\pi}{2l+1}
 \frac{q_{\Lambda}^{c\,(\vec{k},\vec{K},s_z)}(R)}{R^{l+1}} .
 \label{6.21}  
\end{eqnarray}
\end{mathletters}
Then the potential inside the sphere due to the ``exchange charges"
outside is
\begin{eqnarray}
 & & V_{e,S}^{out}(\vec{r})=\frac{1}{N} 
 e^{i \vec{q} \vec{X}(\vec{n})} \,[ \, V_{e,\, 0}^{out} +
 {\sum_{\Lambda \neq 0}}' V_{e,\, \Lambda}^{out}
 \left( \frac{r}{R} \right)^l S_{\Lambda}(\hat{r}) ] . \nonumber \\
 & & \label{6.23} 
\end{eqnarray}
Therefore, Eqs.\ (\ref{6.20b}), (\ref{6.23}) and (\ref{2.21}) where
$q_0(r)$, $q'_0(r)$, $q_{\Lambda}(r)$, $q_{\Lambda'}(r)$ are
given by (\ref{6.21a}-d) fully determine the effective ``exchange"
potential inside any sphere. 

In order to obtain the matrix element of exchange we integrate
the ``exchange" potential with the ``exchange" density
$\rho_{c\,(\vec{k},\vec{K},s_z)}^*(\vec{r})$, Eqs.\ (\ref{6.5})-(\ref{6.7}),
and recall that $c$ stands for $(\vec{p},\beta)$.
From the potentials inside the spheres and in interstices 
we distinguish three contributions to the exchange,
\begin{eqnarray}
 \langle \vec{k}, \vec{P},s'_z | {\cal F}^{exc}(\vec{p},\beta) |
  \vec{k}, \vec{K},s_z \rangle
  =  \frac{1}{N} (B_{out}^{exc}(\vec{p},\beta)
    \nonumber \\
  +B_{I}^{exc}(\vec{p},\beta)+ B_{S}^{exc}(\vec{p},\beta)) .
  \label{6.24} 
\end{eqnarray}
Here $B_{out}^{exc}$, $B_{S}^{exc}$ and $B_{I}^{exc}$  are
the contributions from integrations with $V_{e,S}^{out}(\vec{r})$,
$U_{e,S}(\vec{r})$ inside a sphere, and with $V_e(\vec{R})$
in the interstices, respectively.
First we carry out the integrations
inside a sphere $S(\vec{n})$, and
then perform summation over the $N$ spheres.
As a result we get
\begin{eqnarray}
 B_{out}^{exc}(\vec{p},\beta) =  
 {\sum_{\Lambda \neq 0}}' V_{e,\, \Lambda}^{out}
  \frac{q_{\Lambda}^{c\,(\vec{k},\vec{P},s'_z)}(R)}{R^l} .
  \label{6.25} 
\end{eqnarray}
The contribution from the interstitial region gives
\begin{eqnarray}
 B^{exc}_I(\vec{p},\beta)= \sum_{\vec{K}',\vec{P}}
 V_e(\vec{K}')\, \gamma_{\vec{P}+\vec{K}}(\vec{p},\beta,s'_z)\, 
 {\cal O}(\vec{K}'-\vec{P}) ,
  \label{6.26} 
\end{eqnarray}
where ${\cal O}(\vec{K}'')$ is the overlap integral (\ref{7.4b}).
Finally, for $B^{exc}_S$ we get
\begin{eqnarray}
 & & B^{exc}_S(\vec{p},\beta)   = 
  \sum_{\Lambda} \sum_{l_1 l_2} \sum_{l'_1 l'_2}
 \sum_{\lambda_1 \lambda_2} \sum_{\lambda'_1 \lambda'_2} 
 c_{\Lambda}(\lambda_1,\lambda_2) c_{\Lambda}(\lambda'_1,\lambda'_2) 
 \nonumber \\
 & & \times  
 [\gamma A]_{\lambda_1}^{p\,*}(\vec{p},\beta,s'_z)\,
 [\gamma A]_{\lambda'_1}^{p'}(\vec{p},\beta,s_z)\,
 A_{\lambda_2}^t(\vec{k},\vec{K})\,
 A_{\lambda'_2}^{t'\,*}(\vec{k},\vec{P})
 \nonumber  \\
 & & \times C_l(u_{l_1}^p u_{l_2}^t|u_{l'_1}^{p'} u_{l'_2}^{t'})  .
  \label{6.28}
\end{eqnarray}
Notice that all contributions to the Fock exchange, 
Eqs. (\ref{6.24})-(\ref{6.26}) and (\ref{6.28}) 
are proportional to $1/N$ as one
could expect from a charge distributed over $N$ cells.
From Eq.\ (\ref{6.24}) we can calculate the Coulomb self-energy
associated with a conduction state $| \vec{k}, \alpha \rangle$.
Assuming $\vec{p}=\vec{k}$, $\alpha=\beta$
and taking into account the expansion (\ref{b.5a}) and (\ref{b.6}), 
we arrive at
\begin{eqnarray}
  E_{si}(\vec{k}, \alpha) & = & V_{h}  +  \sum_{\vec{K},\vec{P}} \sum_{s_z,s'_z}
  \gamma_{\vec{P}}^*(\vec{k},\alpha,s'_z)\,
  \gamma_{\vec{K}}(\vec{k},\alpha,s_z)       \nonumber \\
 & \times &
  \langle \vec{k}, \vec{P},s'_z | {\cal F}^{exc}(\vec{k},\alpha) 
  | \vec{k}, \vec{K}, s_z \rangle
  \sim \frac{1}{N} ,
 \label{6.29}
\end{eqnarray}
where $\langle \vec{k}, \vec{P},s'_z | {\cal F}^{exc}(\vec{k},\alpha) 
  | \vec{k}, \vec{K}, s_z \rangle$ is given by (\ref{6.24}).
Here we have introduced 
$V_h$ which is an electrostatic energy, 
associated with the electron charge $e$
distributed homogeneously inside the crystal. (Such homogeneous
charge distribution is absent for exchange 
if two conduction states are different, Appendix F. 
It is compensated by the positive contribution
of nuclei for the direct Coulomb interaction.)
$V_h$ depends on the shape of a crystal, but decreases as
$1/N$. As a result we observe that $E_{si} \rightarrow 0$
in the limit $N \rightarrow \infty$ for any extended state. 

By summing the exchange over all occupied conduction states 
$| \vec{p}, \beta \rangle \neq | \vec{k}, \alpha \rangle$
we obtain a finite value of exchange for each conduction
electron $| \vec{k},\alpha \rangle$,
\begin{eqnarray}
 \langle \vec{k}, \vec{P},s'_z | {\cal F}^{exc}(val) | 
 \vec{k}, \vec{K},s_z \rangle
  =  \frac{1}{N} \sum_{\vec{p},\beta} (B_{out}^{exc}(\vec{p},\beta)
    \nonumber \\
  +B_{I}^{exc}(\vec{p},\beta)+ B_{S}^{exc}(\vec{p},\beta)) .
  \label{6.30} 
\end{eqnarray}
In the absence of the spin-orbit coupling, the matrix elements
(\ref{6.24}) and (\ref{6.30}) are diagonal in spin components $s_z=\pm1/2$.
An important practical complication here arises due to the fact that
the structural constants $V_{e\, 0}$,
$V_{e\, \Lambda}$ ($V_{e\, 0}^{out}$, $V_{e\, \Lambda}^{out}$) 
depend on $\vec{p}$ ($\vec{q}$) and the calculation of them
has to be repeated for each vector $\vec{p}$.

\section {Conclusions} 
\label{sec:cs} 

We have presented a new Hartree-Fock-LAPW method for electron band
structure calculations. The method combines the
restricted Hartree-Fock-Roothaan approach
with the crystalline basis functions in the form of
linear augmented plane waves. 
The strategy of the full potential LAPW treatment \cite{Wei,Rud} is adopted
for calculations of the matrix elements of the direct Coulomb
interactions and exchange.
This is pivotal for collecting all exchange terms together
including  the long-range and multipole contributions.
 
In the framework of the FP-LAPW treatment
an original technique for the solution of periodic Poisson's equation
is formulated, Sec.\ II.
The technique takes into account the partitioning of space into two regions,
inside the spheres and in the interstices, Fig.\ 1. 
In the interstitial region
we expand electron densities and the potential in Fourier series
and express ``exchange" densities in terms of plane waves. 
Inside the spheres we expand densities and
potentials in multipole series.
Finally, we use these expansions to calculate the matrix elements 
of the direct Coulomb interaction (Sec.\ IV) and the exchange (Sec.\ V).
The crystal field effects are considered for core electron shells
and for conduction electrons. These
effects are associated with the nonspherical density components
of $f$, $d$ and, for noncubic symmetries, of $p$
electrons. \cite{Nik1,Nik2,Nik3}
There, the crystal site symmetry is taken into
account and the basis functions are adapted for the spin-orbit interaction.

The technique for solving Poisson's equation has been
applied to the face centered cubic lattice, Sec.\ III.
We have calculated structural constants which are used
to restore cubic Coulomb potentials inside a sphere from its monopole 
($l=0$) and multipole ($l=4,6$) moments. 
We have compared our technique with the pseudo-charge-density method 
of Weinert, Ref.\ \onlinecite{Wei} and the two-center expansion
of the Coulomb interaction, Ref.\ \onlinecite{Nik1}.

At present we are working on programming the formulas derived
in this article. However, it is already clear that the task
consists of two independent parts. First of all, one should calculate
the multipole matrix elements $c_{\Lambda}(\lambda_1,\lambda_2)$
of electron transitions, Eq.\ (\ref{3.20}), and 
the other coefficients related to them 
(such as $c_{\Lambda}(\tau,\tau')$, Eq.\ (\ref{3.27}), and 
$c_{\Lambda}(\tau,s_z;\, \lambda)$, Eq.\ (\ref{5.7b})).
The integrations there involve only angular (and spin) parts
of electronic wave functions and thus the coefficients can be tabulated 
and stored before the HFR self-consistent procedure.
Also to this part one should add calculations of the relevant structural
constants, such as $V_{40}$, $V_{44}^{out}$, $V_{46}$, $V_{60}$, and
$V_{66}^{out}$ computed in Sec.\ III for the face centered cubic
structure. 
(The constants are needed to restore the full potential for
a given set of multipole moments.)
These calculations depend on the type of crystal symmetry
but are separated from the problems of HFR method.
The second task is to program the matrix elements and all
relevant procedures of the restricted HFR method. 
For those purposes one can start with an existing code of LAPW method
and develop it on the basis of the considerations presented
in this article.


\acknowledgments 
We thank 
professor K.H. Michel for numerous fruitful discussions and
S. Balaban, O. Kepp and D. Kirin for helpful remarks.
This work has been financially supported by the Fonds voor
Wetenschappelijk Onderzoek, Vlaanderen, and by the Russian
Foundation for Basic Reasearch, project No. 00-03-32968.

\appendix

\section{} 
\label{sec:apA}

Throughout this paper instead of complex (surface) spherical
harmonics $Y_l^m$ we use real symmetry adapted functions \cite{Bra} (SAFs)  
$S_{\Lambda}$, which
transform according to irreducible representations $\Lambda(l)$
of a site symmetry group $G$. The composite index
$\Lambda(l)$ stands for $(l;\Gamma(l), \mu(l), k)$ where $\Gamma(l)$
labels the irreducible representations within the $l$ manifold,
$\mu(l)$ numbers the representations that occur more than once
and $k$ denotes the rows of a given representation.
In the following we omit index $l$ in $\Gamma$ and $\Lambda$.
SAFs $S_{\Lambda}$ are linear combinations of $Y_l^m$ with
the same $l$, {\it i.e},
\begin{eqnarray}
  S_{\Lambda}=\sum_{m=0}^l 
   c^{\Lambda}_{m,c} Y_l^{m,c} + 
  \sum_{m=1}^l c^{\Lambda}_{m,s} Y_l^{m,s}  ,
 \label{a.1}
\end{eqnarray}
where the coefficients $c^{\Lambda}_{m,c}$ and
$c^{\Lambda}_{m,s}$ depend on the group $G$
under consideration,  and
$Y_l^{m=0,c} \equiv Y_l^0$.
The coefficients $c^{\Lambda}_{m,c,s}$ for different groups 
are quoted in Tables 2.4-2.6 of Ref.~\onlinecite{Bra}.
There are $(2l+1)$ independent SAFs $S_{\Lambda}$ belonging
to the $l$ manifold.
The real spherical harmonics are  
\begin{mathletters}
\begin{eqnarray}
 & & Y_l^{m,c}=\frac{1}{\sqrt{2}}(Y_l^m+Y_l^{-m})  \label{a.2a} \\
 & & Y_l^{m,s}=\frac{1}{i\sqrt{2}}(Y_l^m-Y_l^{-m}) \label{a.2b}  
\end{eqnarray}
\end{mathletters}
where $Y_l^m$ are taken with the phase definition of Ref.~\onlinecite{Bra}.
(It is different from the definition used by Condon and Shortly. \cite{CS})
For a given $l$  in a $(2l+1)$ dimensional space
we consider row vectors $(Y_l^{0,c},Y_l^{1,c},...,
Y_l^{l,c},Y_l^{1,s},...,Y_l^{l,s}) \equiv Y$ (we exclude $Y_l^{0,s} \equiv 0$),
and $(S_{\Lambda}) \equiv S$, and the matrix
$\left[{}^{c^{\Lambda}_{m,c}}_{c^{\Lambda}_{m,s}} \right] 
\equiv c$.
Then the SAFs and the spherical harmonics for a given $l$ 
are connected through an orthogonal transformation,
\begin{eqnarray}
 S=Y \cdot c . \label{a.3}
\end{eqnarray}
One can easily find the inverse transformation since
\begin{eqnarray}
 Y=S \cdot c^T ,  \label{a.4}
\end{eqnarray}
where $T$ stands for the transpose, since $c^{-1}=c^T$.
It is more convenient to use $S_{\Lambda}$ than $Y_l^m$ due
to their known symmetry properties.

In the density expansion, Eq.\ (\ref{2.2b}), only the SAFs of $A_{1g}$ symmetry
survive because density stays invariant under all symmetry
operations of $G$. 
However, for calculations of exchange (Sec.\ V) there is no such
simplification and
the full basis set (including SAFs belonging to the
other irreducible representations) should be taken into account. 

We use SAFs to describe both electronic densities and 
wave functions.  
For electronic states we adopt a notation with
small letters, {\it i.e} $\lambda=(l;\Gamma,\nu,k)$.
For localized electrons for conciseness we incorporate also the
principal quantum number $n$ and write
$\tau=(n,l;\Gamma,\nu,k)$.
In the latter case $\Gamma$
refers to a double valued irreducible representations of
$G$. \cite{Bra}

\section{} 
\label{sec:apB}

Here we introduce some definitions and notations 
of the linear augmented plane wave method
(LAPW). \cite{And,Koe,Sin} 
The coordinate basis functions are plane waves in the
interstitial region,
\begin{eqnarray}
 \chi_{\vec{k},\vec{K}}(\vec{R}) \equiv 
 \langle \vec{R} | \vec{k}, \vec{K} \rangle =
  \frac{1}{\sqrt{N v}} 
   e^{i (\vec{k}+\vec{K}) \cdot \vec{R} } ,
 \label{b.1}
\end{eqnarray}
and a linear combination of local atomic functions
inside the spheres,
\begin{eqnarray}
& & \langle \vec{R} \in S | \vec{k}, \vec{K} \rangle 
 = \frac{1}{\sqrt{N}} \sum_{\vec{n}} e^{i \vec{k} \cdot \vec{X}(\vec{n})}
 \sum_{\lambda}  
 \phi_{\lambda}^{\vec{k}, \vec{K}}(\vec{R}-\vec{X}(\vec{n})) ,
 \nonumber \\
 & & \label{b.2}
\end{eqnarray}
where \cite{Koe}
\begin{eqnarray}
 & & \phi_{\lambda}^{\vec{k}, \vec{K}}(\vec{r})= \left(
  A_{\lambda}(\vec{k},\vec{K})\, u_{l}(r)+
  B_{\lambda}(\vec{k},\vec{K})\, \dot{u}_{l}(r) \right) S_{\lambda}(\hat{r}) .
  \nonumber \\
 & & \label{b.3}
\end{eqnarray}
(If one uses spherical harmonics $Y_l^m$ instead of SAFs $S_{\lambda}$,
then $\lambda=(l,m)$.)
In order to condense notations we introduce two components ($p$=1,2)
of the radial function $u_l^p(r)$,
\begin{eqnarray}
  u_l^1(r) \equiv u_l(r),   \;\;\;\;   
  u_l^2(r) \equiv \dot{u}_l(r)=\frac{\partial u_l(r,E)}{\partial E} .
 \label{b.4} 
\end{eqnarray}
and the corresponding to them two components $A_{\lambda}^p(\vec{k},\vec{K})$, 
which are
\begin{eqnarray}
  A_{\lambda}^1(\vec{k},\vec{K}) \equiv A_{\lambda}(\vec{k},\vec{K}),
  \;\;\;\;   
  A_{\lambda}^2(\vec{k},\vec{K}) \equiv B_{\lambda}(\vec{k},\vec{K})  .
 \label{b.5} 
\end{eqnarray}
The coefficients $A_{\lambda}^p$ are obtained by requiring that
the basis functions and their derivatives are continuous on
the sphere boundary. \cite{Koe}

In the absence of a static magnetic field
and the spin-orbit coupling,
the conduction electronic states with spin projections $s_z=\pm 1/2$
are degenerate. (The spin-orbit coupling can be included later in
the second variational treatment as described in Ref.\ \onlinecite{Sin}.)
The wave function of a conduction electron with the wave vector $\vec{k}$ 
and the band index $\alpha$ then reads \cite{Eli,Tin}
\begin{eqnarray}
   \langle \vec{R} | \vec{k}, \alpha \rangle = \sum_{s_z=-1/2}^{1/2}
 \langle \vec{R},s_z | \vec{k}, \alpha \rangle \zeta(s_z) ,
 \label{b.5a} 
\end{eqnarray}
where $\zeta(s_z=\pm1/2)$ are the two spinors and
\begin{eqnarray}
   \langle \vec{R},s_z | \vec{k}, \alpha \rangle =
   \sum_{\vec{K}} \gamma_{\vec{K}}(\vec{k},\alpha,s_z) 
   \langle \vec{R} | \vec{k}, \vec{K} \rangle .
 \label{b.6} 
\end{eqnarray}
The coefficients $\gamma_{\vec{K}}(\vec{k},\alpha,s_z)$
are found by the HF variational procedure. \cite{Sin}
Inside a sphere $S(\vec{n})$ the wave function 
is given by
\begin{eqnarray}
  & & \langle \vec{r},s_z | \vec{k}, \alpha \rangle 
   = \frac{e^{i \vec{k} \cdot \vec{X}(\vec{n})}}{\sqrt{N}}
   \sum_{\lambda}  
   [\gamma A]_{\lambda}^p(\vec{k},\alpha,s_z) \, 
   u_l^p(r)\, S_{\lambda}(\hat{r}) , \nonumber \\
 & & \label{b.7} 
\end{eqnarray}
where $\vec{r}=\vec{r}(\vec{n})$ and summation over $p=1,2$ is implied. 
Here we have introduced the notation
\begin{eqnarray}
   [\gamma A]_{\lambda}^p(\vec{k},\alpha,s_z) 
   = \sum_{\vec{K}} \gamma_{\vec{K}}(\vec{k},\alpha,s_z)
   A_{\lambda}^p(\vec{k},\vec{K}) .
 \label{b.8} 
\end{eqnarray}

\section{} 
\label{sec:apA1}

Here we derive analytical orientational
wave vectors for the cubic site symmetry $O_h$
by employing the eigenvectors tabulated in Ref.~\onlinecite{Lea}.

The core states of $s_{1/2}$, $p_{1/2}$, $p_{3/2}$ and $d_{3/2}$ electrons 
remain degenerate while $d_{5/2}$, $f_{5/2}$ and $f_{7/2}$ 
are split in the cubic environment.
If $D_{5/2}$ and $D_{7/2}$ are doubled valued
representations of $SO(3)$, then the symmetry lowering is \cite{Tin}
\begin{mathletters}
\begin{eqnarray}
 & & D_{5/2} \rightarrow \Gamma_7 + \Gamma_8 , \label{a.5a} \\
 & & D_{7/2} \rightarrow \Gamma_7 + \Gamma_8 + \Gamma_6 . \label{a.5b}
\end{eqnarray}
\end{mathletters}
For two components of the doublet $\Gamma_7$ of $d_{5/2}$ we
have found
\begin{mathletters}
\begin{eqnarray}
 \langle \hat{r} | d_{5/2}, (\Gamma_7,1) \rangle &=&
 \frac{i}{\sqrt{3}} S_{2,(T2g,3)} \zeta_+  \nonumber \\
 & &+\frac{1}{\sqrt{3}}(iS_{2,(T2g,1)}-S_{2,(T2g,2)}) \zeta_-, \label{a.6a} \\
 \langle \hat{r} | d_{5/2}, (\Gamma_7,2) \rangle &=&  
 \frac{1}{\sqrt{3}}(iS_{2,(T2g,1)}+S_{2,(T2g,2)}) \zeta_+ \nonumber \\
 & &-\frac{i}{\sqrt{3}} S_{2,(T2g,3)} \zeta_-, \label{a.6b}
\end{eqnarray}
\end{mathletters}
where $S_{2,(T2g,k)}$ refer to the three components ($k=1-3$) of $T_{2g}$ 
symmetry ($l=2$) given in Table 2.6 of Ref.\ \onlinecite{Bra}.
The second component, Eq.\ (\ref{a.6b}), is connected with the first, 
Eq.\ (\ref{a.6a}), through the time reversal symmetry.
One can easily check it by noting
that the SAFs are real and by applying the following rules for 
the time reversal symmetry
\begin{eqnarray}
 i \rightarrow -i , \;\;\; \zeta_+ \rightarrow \zeta_- , 
 \;\;\;  \zeta_- \rightarrow -\zeta_+ . \label{a.7}
\end{eqnarray}
For two components of $\Gamma_8$ we have 
\begin{mathletters}
\begin{eqnarray}
 \langle \hat{r} | d_{5/2}, (\Gamma_8,1) \rangle &=&
 \frac{1}{\sqrt{15}}(i 2 S_{2,(T2g,3)} +3S_{2,(Eg,2)}) \zeta_+  \nonumber \\
  &+& \frac{1}{\sqrt{15}}(-iS_{2,(T2g,1)}+S_{2,(T2g,2)}) \zeta_-, \label{a.8a} \\
 \langle \hat{r} | d_{5/2}, (\Gamma_8,2) \rangle &=&  
 \sqrt{\frac{3}{5}} S_{2,(Eg,1)} \zeta_+ \nonumber \\
 &-& \frac{1}{\sqrt{5}}(iS_{2,(T2g,1)}+S_{2,(T2g,2)}) \zeta_- . \label{a.8b}
\end{eqnarray}
\end{mathletters}
The other two components are obtained from (\ref{a.8a},b) 
by employing the rules (\ref{a.7}).

We consider analogously the splitting of the $f_{5/2}$ ($l=3$) states 
in the cubic symmetry and obtain
\begin{eqnarray}
 \langle \hat{r} | f_{5/2}, (\Gamma_7,1) \rangle &=&
 \frac{1}{\sqrt{7}}(-\sqrt{3}i S_{3,A2u}+\frac{2}{\sqrt{3}}S_{3,(T2u,3)} ) 
 \zeta_+  \nonumber \\
  &+& \frac{2}{\sqrt{21}}(S_{3,(T2u,1)}+iS_{3,(T2u,2)}) \zeta_-, 
 \label{a.9} 
\end{eqnarray}
and
\begin{mathletters}
\begin{eqnarray}
 & & \langle \hat{r} | f_{5/2}, (\Gamma_8,1) \rangle =
 -\sqrt{\frac{5}{21}} S_{3,(T2u,3)} \zeta_+ 
 +[ \frac{3}{2\sqrt{7}} (S_{3,(T1u,1)} \nonumber \\
 & & -iS_{3,(T1u,2)}) 
 +\frac{1}{2}\sqrt{\frac{5}{21}} 
 (S_{3,(T2u,1)}+iS_{3,(T2u,2)})] \zeta_-, 
 \label{a.10a} \\
 & & \langle \hat{r} | f_{5/2}, (\Gamma_8,2) \rangle =
 \sqrt{\frac{3}{7}} S_{3,(T1u,3)} \zeta_+ 
 -\frac{1}{2\sqrt{7}}
 [ \sqrt{3} (S_{3,(T1u,1)} \nonumber \\
 & & +iS_{3,(T1u,2)}) 
 - \sqrt{5} 
 (S_{3,(T2u,1)}-i S_{3,(T2u,2)})] \zeta_- . 
 \label{a.10b} 
\end{eqnarray}
\end{mathletters}
Real SAFs $S_{3,A2u}$, $S_{3,(T1u,k)}$ and $S_{3,(T2u,k)}$
($k=1-3$) are given in Table 2.6 of Ref.~\onlinecite{Bra}.
The other components are obtained through the time reversal symmetry,
Eq.\ (\ref{a.7}).

Finally, for $f_{7/2}$ core states we get
\begin{mathletters}
\begin{eqnarray}
 & & \langle \hat{r} | f_{7/2}, (\Gamma_6,1) \rangle =
  \frac{1}{\sqrt{3}} ( S_{3,(T1u,1)}-iS_{3,(T1u,2)}) \zeta_+ 
 \nonumber \\
 & & - \frac{1}{\sqrt{3}} 
 S_{3,(T1u,3)} \zeta_-, 
 \label{a.11a} \\
 & & \langle \hat{r} | f_{7/2}, (\Gamma_7,1) \rangle =
  \frac{1}{\sqrt{7}}(2i S_{3,A2u}+S_{3,(T2u,3)}) \zeta_+ 
  \nonumber \\
 & & + \frac{1}{\sqrt{7}} 
 (S_{3,(T2u,1)}+i S_{3,(T2u,2)}) \zeta_- . 
 \label{a.11b} 
\end{eqnarray}
\end{mathletters}
These are first components of $\Gamma_6$ and $\Gamma_7$, respectively. 
For the quartet $\Gamma_8$ of $f_{7/2}$ we have found
\begin{mathletters}
\begin{eqnarray}
 & & \langle \hat{r} | f_{7/2}, (\Gamma_8,1) \rangle =
  \frac{1}{2}[ \sqrt{\frac{5}{21}} 
  ( S_{3,(T1u,1)}-iS_{3,(T1u,2)}) 
 \nonumber \\
  & & + \frac{3}{\sqrt{7}} ( S_{3,(T2u,1)}+iS_{3,(T2u,2)}) ] \zeta_+ 
 +\sqrt{\frac{5}{21}} S_{3,(T1u,3)} \zeta_-, 
 \label{a.12a} \\
 & & \langle \hat{r} | f_{7/2}, (\Gamma_8,2) \rangle =
  \sqrt{\frac{3}{7}} S_{3,(T2u,3)} \zeta_+ 
  +\frac{1}{2} [ \sqrt{\frac{5}{7}}(S_{3,(T1u,1)}
  \nonumber \\
 & & -iS_{3,(T1u,2)}) - \sqrt{\frac{3}{7}} 
 (S_{3,(T2u,1)}+i S_{3,(T2u,2)}) ] \zeta_-  . 
 \label{a.12b} 
\end{eqnarray}
\end{mathletters}
Again, the other components of $\Gamma_6$, $\Gamma_7$ and
$\Gamma_8$, are found by applying the time reversal
transformation (\ref{a.7}) to (\ref{a.11a},b)
and (\ref{a.12a},b).

\section{} 
\label{sec:apE}

Here we describe a simple method to generate
radial basis functions.
We can use the spherically symmetric component of
the total electrostatic potential $V_0(\vec{r})$
(inside a sphere $S(\vec{n})$)
to find the radial solution $u_{n,l}$ and the 
corresponding energy $E_{n,l}$,
\begin{eqnarray}
    h_{l} u_{n,l} = E_{n,l} u_{n,l} .
 \label{e.1}
\end{eqnarray}
Here $n$ is the principal quantum number,
and $h_{l}$ is the radial operator of
the Schr\"{o}dinger equation (Eq.\ (1b) of Ref.~\onlinecite{Koe}).
We assume that the solutions are confined inside the
sphere $S(\vec{n})$ and
on the sphere boundary $r=R$ we have
\begin{eqnarray}
    u_{l}(R)  = 0 ,\;\;\;\; 
    \left.  \frac{\partial u_{l}(r)}{\partial r} \right|_R = 0.
 \label{e.3}
\end{eqnarray}
The boundary conditions are complementary to (\ref{e.1}).
Starting with (\ref{e.1}) and (\ref{e.3}) one obtains $u_{n,l}$ and $E_{n,l}$.

For a given $l$ we consider the radial functions $u_{n,l}$
which differ from each other by the principal quantum number $n=l+1,l+2,...$~.
These functions correspond to the same angular dependence 
(specified by $l,m_l$) and
form a complete orthonormalized set. 
Therefore, they can serve as a basis for
any radial function ${\cal R}_{\tau}$ 
satisfying Eq.\ (\ref{e.3}). A nonrelativistic
function of a localized electron is characterized
by the combined index $\tau \equiv (l,m_l)$ and we write
$u_{\eta}^{\tau}=u_{n,l}$, $i.e.$ $\eta=n$.
For the relativistic case one has to distinguish two solutions
with the total momentum $j=l-1/2$ and $j=l+1/2$.
Then $\tau=(l,j,m_j)$ and the basis functions are
$u_{\eta}^{\tau}$, where again $\eta=n$.  

For the matrix elements of the spherically symmetric potential $V_0^{\tau}$,
 Eq.\ (\ref{4.4}), we find
\begin{eqnarray}
  \langle \tau, \eta | {\cal F}_0^{Coul} | \tau, \eta' \rangle & = & 
  E_{n,J}  \delta_{\eta \eta'}
  - \frac{1}{4\pi}
  C_0(u_{\eta}^{\tau} u_{\eta'}^{\tau}| {\cal R}_{\tau} {\cal R}_{\tau}) . 
 \label{e.7}
\end{eqnarray}
Here $E_{n,J}$ is the eigenvalue corresponding to $u_{\eta}^{\tau}$
($J=(l,j)$)
and the integral $C_0(...)$ is given by Eq.\ (\ref{d.1}).

\section{} 
\label{sec:apD}

The two-fold integral $C_l(f_1,f_2|g_1 g_2)$ 
of four radial functions $f_1(r)$, $f_2(r)$, $g_1(r)$ and $g_2(r)$
is defined as
\begin{eqnarray}
 C_l(f_1,f_2|g_1 g_2)   & = & 
 \int_0^R dr\, r^2 \int_0^R dr'\, {r'}^2 \nonumber \\
 & & \times f_1(r) f_2(r)\, v_{l}(r,r') \, g_1(r') g_2(r')
 \label{d.1}
 \end{eqnarray}
where the one center multipole
function is given by
$v_{l}(r,r') =\frac{4\pi}{2l+1} \frac{r_<^l}{r_>^{l+1}}$, 
Eq.\ (\ref{3.16d}), and $r_<$ $(r_>)$ is the smaller (larger) of $r$ and $r'$.

\section{} 
\label{sec:apF}

We consider the exchange between two extended states 
$| d \rangle = | \vec{k}, \alpha \rangle$
and $| c \rangle = | \vec{p}, \beta \rangle$.
We expand both states in the LAPW basis functions, Eq.\ (\ref{b.6}).
Proceeding as in section V.C, we introduce an effective ``exchange" density
$\rho_{cd}(\vec{R})$, Eq.\ (\ref{5.4}),
and the corresponding ``exchange charge":
\begin{eqnarray}
  Q^{c\,d}_0(R) &=& \sum_{l} \sum_{\lambda(l)} \sum_{s_z}
 [\gamma A]_{\lambda}^{p\, *}(\vec{p},\beta,s_z)\, 
 [\gamma A]_{\lambda}^p(\vec{k},\alpha,s_z)\,
 N_l^p  \nonumber \\
 & & +  {\cal O}(\vec{p},\beta;\vec{k},\alpha),
 \label{f.1} 
\end{eqnarray}
where
\begin{eqnarray}
& & {\cal O}(\vec{p},\beta;\vec{k},\alpha) =
 - \frac{4\pi R^2}{v}
 \sum_{\vec{K}'} \frac{j_1(|\vec{K}'+\vec{q}|R)}{|\vec{K}'+\vec{q}|}
  \rho_{\vec{K}'}(\vec{p},\beta;\vec{k},\alpha)   \nonumber \\
& & \label{f.2}
\end{eqnarray}
and
\begin{eqnarray}
 \rho_{\vec{K}'}(\vec{p},\beta;\vec{k},\alpha)=\sum_{\vec{K}} \sum_{s_z}
  \gamma^*_{\vec{K}'+\vec{K}}(\vec{p},\beta,s_z)
  \gamma_{\vec{K}}(\vec{k},\alpha,s_z) .
 \label{f.3} 
\end{eqnarray}
On the other hand,
the orthogonality relation for the two extended states is
\begin{eqnarray}
 & & \langle \vec{p},\beta | \vec{k},\alpha \rangle = 
 \delta_{\alpha \beta}  \delta_{\vec{p} \vec{k}}  \nonumber \\
 & & = \sum_{l} \sum_{\lambda(l)} \sum_{s_z} 
 [\gamma A]_{\lambda}^{p\, *}(\vec{p},\beta,s_z)\, 
 [\gamma A]_{\lambda}^p(\vec{k},\alpha,s_z)\, N_l^p  \nonumber \\ 
 & & +{\cal O}(\vec{p},\beta;\vec{k},\alpha)
 + \delta_{\vec{p} \vec{k}} 
 \sum_{\vec{K}} \sum_{s_z} \gamma_{\vec{K}}^*(\vec{p},\beta,s_z)\,
 \gamma_{\vec{K}}(\vec{k},\alpha,s_z) .
 \label{f.4}
\end{eqnarray}
Comparing it with (\ref{f.1}) we observe that if $\vec{p} \neq \vec{q}$
then $Q^{c\,d}_0(R)=0$.
However, it will not be erroneous to use this charges as they appear
in HFR procedure, Eq.\ (\ref{6.12'}). Since Poisson's equation is linear,
their contributions will cancel in the final results.
Generalizing the orthonormality relation (\ref{f.4}) 
for the case of few atoms
one can show that it leads to
\begin{eqnarray}
   \sum_i Q^{cd}_i =0 , \;\;\;\;  \vec{p} \neq \vec{k} .
 \label{f.5}
\end{eqnarray}
We conclude that the ``exchange" charges $Q^{cd}_i$ are not necessarily zero.
(Here $i$ labels different
atoms in the unit cell.)
The orthogonality relation (\ref{f.5}) ensures that
there is no Coulomb divergence associated with the uniform component
of the ``exchange density".




\begin{references} 

\vspace{-1cm}
 

\bibitem{DFT}
R.M. Dreizler and E.K.U. Gross, {\it Density Functional Theory:
An approach to the quantum many-body problem} (Springer, Berlin, 1990);
W. Kohn and P. Vashishta, in {\it Theory of the inhomogeneous electron
gas}, S. Lundqvist and N.H. March, Eds., (Plenum, New York, 1983), p. 79.
\bibitem{Arg}
R.O. Jones and O. Gunnarsson, Rev. Mod. Phys. {\bf 61}, 689 (1989);
N. Argaman and G. Makov, cond-mat/9806013.
\bibitem{GGA}
D.C. Langreth and J.P. Perdew, Phys. Rev. B {\bf 21}, 5469 (1980);
J.P. Perdew and Y. Wang, Phys. Rev. B {\bf 33}, 8800 (1986);
J.P. Perdew J.A. Chevary, S.H. Vosko, K.A. Jackson, M.R. Pederson, D.J. Singh,
C. Fiolhais, Phys. Rev. B {\bf 46}, 6671 (1992).
%
\bibitem{Nes}
R.K. Nesbet, R. Colle, J. Math. Chem. {\bf 26}, 233 (1999).
\bibitem{SIC}
J.P. Perdew and A. Zunger, Phys. Rev. B {\bf 23}, 5048 (1981).
\bibitem{HFR}
C. Pisani, R. Dovesi and C. Roetti, {\it Hartree-Fock ab-initio
treatment of crystalline systems}, Lecture Notes in Chemistry,
vol. 48, Springer Verlag, Heidelberg, 1988.
\bibitem{Pis}
C. Pisani, R. Dovesi, Int. J. Quantum Chem. {\bf 17}, 501 (1980);
R. Dovesi, C. Roetti, {\it ibid.} {\bf 17}, 517 (1980).
\bibitem{CRY}
V.R. Saunders, R. Dovesi, C. Roetti, M. Caus\`{a}, N.M. Harrison,
R. Orlando, C.M. Zicovich-Wilson, CRYSTAL 98 User's Manual,
Theoretical Chemistry Group, University of Torino (1998).
\bibitem{Sau}
V. R. Saunders, C. Freyria-Fava, R. Dovesi, L. Salasco, C. Roetti,
Mol. Phys. {\bf 77}, 629 (1992).
\bibitem{Dov}
R. Dovesi, C. Pisani, C. Roetti, V.R. Saunders,
Phys. Rev. B {\bf 28}, 5781 (1983).
%
\bibitem{Kud}
for a DFT version, see
K.N. Kudin and G.E. Scuseria, Phys. Rev. B {\bf 61}, 16440 (2000);
a HF version is quoted in P.Y. Ayala, K.N. Kudin, G.E. Scuseria,
J. Chem. Phys. {\bf 115}, 9698 (2001).
%
\bibitem{Andre}
polymer code PLH93, see J.M. Andr\'e, D.H. Mosley, B. Champagne, J. Delhalle,
J.G. Fripiat, J.L. Br\'edas, D.J. Vanderveken, and D.P. Vercauteren, in
{\it METECC-94, Methods and Techniques in Computational Chemistry},
edited by E. Clementi (STEF, Caligari, 1993), v. B, Chap. 10, p. 423. 
%
\bibitem{Vis}
MOLFDIR, see L. Visscher, O. Visser, P.J.C. Aerts, H. Merenga,
W.C. Nieuwpoort, Comput. Phys. Commun. {\bf 81}, 120 (1994);
A. Hu, P. Otto, J. Ladik, Chem. Phys. Lett. {\bf 293}, 277 (1998).
\bibitem{And}
O.K. Andersen, Phys. Rev. B {\bf 12}, 3060 (1975).
\bibitem{Koe}
D.D. Koelling and G.O. Arbman, J. Phys. F {\bf 5}, 2041 (1975).
\bibitem{Sin}
D.J. Singh, {\it Planewaves, Pseudopotentials and the LAPW
method}, (Kluwer, Boston, 1994).
\bibitem{Roo}
C.C. Roothaan, Rev. Mod. Phys. {\bf 23}, 69 (1951).
\bibitem{Wei} 
 M. Weinert, J. Math. Phys. {\bf 22}, 2433 (1981).
\bibitem{Rud}
W.E. Rudge, Phys. Rev. {\bf 181}, 1020 (1969).
\bibitem{Ewa}
P.P. Ewald, Ann. Physik {\bf 64}, 253 (1921),
M.P. Tosi, Solid State Phys. {\bf 16}, 1 (1964).
\bibitem{Ewa1}
J.M. Ziman,
{\it Principles of the theory of solids},
(University Press, Cambridge, 1972), p. 39.
\bibitem{Nik1}
A.V. Nikolaev and K.H. Michel, Eur. Phys. J. B {\bf 9},
619 (1999); {\bf 17}, 363 (2000).
\bibitem{Nik2}
A.V. Nikolaev and K.H. Michel, Eur. Phys. J. B,
{\bf 17}, 15 (2000).
\bibitem{Nik3}
A.V. Nikolaev and K.H. Michel, Phys. Rev. B
{\bf 63}, 104105 (2001).
\bibitem{Hut}
M.T. Hutchings, in {\it Solid State Physics: Advances in Research
and Applications}, Eds. F. Seitz and D. Turnbull, v. 16,
(Academic Press, New York, 1964), p.~227.
\bibitem{Wim}
E. Wimmer, H. Krakauer, M. Weinert, A.J. Freeman,
Phys. Rev. B {\bf 24}, 864 (1981).
\bibitem{Ham}
D.R. Hamann, Phys. Rev. Lett. {\bf 42}, 662 (1979).
\bibitem{Ely}
 N. Elyashar and D.D. Koelling, Phys. Rev. B {\bf 13}, 5362 (1976);
 Phys. Rev. B {\bf 15}, 3620 (1977). 
\bibitem{Bra} 
 C.~J.~Bradley and A.~P.~Cracknell, 
 {\it The Mathematical Theory of Symmetry in Solids},
 (Clarendon, Oxford, 1972).
\bibitem{Jac}
D.D. Jackson, {\it Classical Electrodynamics},
(Wiley, New York, 1962).
\bibitem{Abr}
M. Abramowitz and I.A. Stegun, editors, {\it Handbook of
Mathematical Functions} (Dover, New York, 1972).
\bibitem{Hei}
H. Yasuda and T. Yamamoto, Prog. Theor. Phys. {\bf 45}, 1458 (1971);
R. Heid, Phys. Rev. B {\bf 47}, 15912 (1993).
\bibitem{rem}
Explicit expressions for $K_4$ and $K_6$ are given in a number
of publications. See for example Eq.\ (A.8) of Ref.~\onlinecite{Nik1}
and Eq.\ (2.7) of Ref.~\onlinecite{Nik3}.
\bibitem{Eli}
R.J. Elliott, Phys. Rev. {\bf 96}, 266 (1954), {\it ibid}, 280 (1954).
\bibitem{New}
D.J. Newman, Adv. Phys. {\bf 20}, 197 (1971);
D.J. Newman, J. Phys. F: Met. Phys. {\bf 13}, 1511 (1983).
\bibitem{Ste}
L. Steinbeck, M. Richter, U. Nitzsche, H. Eschrig,
Phys. Rev. B {\bf 53}, 7111 (1996).
\bibitem{Bro}
M.S.S Brooks, O. Eriksson, J.M. Wills, B. Johansson,
Phys. Rev. Lett. {\bf 79}, 2546 (1997).
\bibitem{Lea}
K.R. Lea, M.J.M. Leask, W.P. Wolf, J. Phys. Chem. Solids
{\bf 23}, 1381 (1962).
\bibitem{Tin}
M. Tinkham, {\it Group Theory and Quantum Mechanics}
(McGraw-Hill, New York, 1964).
\bibitem{CS}
E.U. Condon and G.H. Shortley,
{\it The theory of atomic spectra}, (University Press, Cambridge, 1967).

\end{references}
\end{document}